\documentclass[aps,prd,reprint,preprintnumbers,nofootinbib,superscriptaddress,floatfix]{revtex4-1}      

\usepackage[normalem]{ulem}        
\usepackage{caption}
\usepackage{tikz,xcolor} 
\usepackage[colorlinks = true,
linkcolor = blue,
urlcolor  = blue,
citecolor = blue,
anchorcolor = blue]{hyperref}
\definecolor{lime}{HTML}{A6CE39}
\DeclareRobustCommand{\orcidicon}{%
	\begin{tikzpicture}
		\draw[lime, fill=lime] (0,0)
		circle [radius=0.16]
		node[white] {{\fontfamily{qag}\selectfont \tiny ID}};
		\draw[white, fill=white] (-0.0625,0.095)
		circle [radius=0.007];
	\end{tikzpicture}
	\hspace{-2mm}
}
\foreach \x in {A, ..., Z}{%
\expandafter\xdef\csname orcid\x\endcsname{\noexpand\href{https://orcid.org/\csname orcidauthor\x\endcsname}{\noexpand\orcidicon}}
}
\usepackage{amsmath}  
\usepackage{amssymb} 
\usepackage{mathtools} 
\usepackage[utf8]{inputenc}

\usepackage{xcolor}   

\usepackage{graphicx,bm}    
\usepackage{booktabs}    
\usepackage{url}

\begin{document} 
%---------------------------------------------------------------------

\title{Revisiting Unidentified Charged-Hadron Fragmentation Functions with Modern COMPASS SIDIS Multiplicities}

\collaboration{HAPS Collaboration}

\author{Maryam Soleymaninia \orcidB{}} 
\email{Maryam\_Soleymaninia@ipm.ir}
\affiliation{School of Particles and Accelerators, Institute for Research in Fundamental Sciences (IPM), Tehran, Iran}

\author{Hamzeh Khanpour \orcidE{}}
\email{Hamzeh.Khanpour@cern.ch}
\affiliation{AGH University, Faculty of Physics and Applied Computer Science, Al. Mickiewicza 30, 30-055 Krakow, Poland}
\affiliation{School of Particles and Accelerators, Institute for Research in Fundamental Sciences (IPM), Tehran, Iran}

\author{Hubert Spiesberger \orcidS{}} 
\email{spiesber@uni-mainz.de}
\affiliation{PRISMA$^{++}$ Cluster of Excellence, Institut f\"ur Physik, Johannes Gutenberg-Universit\"at Mainz, Mainz, Germany}

\author{Majid Azizi \orcidD{}}
\email{Ma.Azizi@ipm.ir}
\affiliation{School of Particles and Accelerators, Institute for Research in Fundamental Sciences (IPM), Tehran, Iran}

\author{Michael~Klasen \orcidG{}}
\email{Michael.Klasen@uni-muenster.de}
\affiliation{Institut f\"ur Theoretische Physik, Universit\"at M\"unster, \\ Wilhelm-Klemm-Stra\ss{}e 9, 48149 M\"unster, Germany}

\author{Hadi Hashamipour \orcidA{}}
\email{H\_Hashamipour@ipm.ir}
\affiliation{School of Particles and Accelerators, Institute for Research in Fundamental Sciences (IPM), Tehran, Iran}

\date{\today}

%---------------------------------------------------------------------
\begin{abstract}  
%---------------------------------------------------------------------

We present \texttt{HAPS-hFF1.0}, a new global QCD analysis of
unidentified charged-hadron fragmentation functions (FFs) using
single-inclusive electron-positron annihilation (SIA) data together with the
modern COMPASS semi-inclusive deep-inelastic scattering (SIDIS)
multiplicities.   
The COMPASS input consists of the 2025 proton-target measurement and
the revised $^6$LiD-target multiplicities provided in the 2026 COMPASS addendum.
The extraction is performed at both next-to-leading order (NLO) and
next-to-next-to-leading order (NNLO), allowing us to study the perturbative
stability of the QCD fit and the impact of the updated SIDIS information on
the flavor structure of the FFs. 
The FFs are parametrized using neural networks, while experimental
uncertainties are propagated through a Monte Carlo replica methodology.
We find that the modern COMPASS multiplicities can be consistently described together with
the SIA data and provide important charge-separated constraints on the
light-quark and antiquark FFs. The comparison between
the NLO and NNLO extractions indicates a stable quark-sector determination,
while the gluon FF remains less directly constrained
in the present SIA+SIDIS analysis. Our results highlight the importance of
the modern COMPASS SIDIS multiplicities for precision studies of
unidentified charged-hadron fragmentation and for future global FF
determinations. 
The resulting \texttt{HAPS-hFF1.0} replicas are publicly available in standard  
LHAPDF format.

%---------------------------------------------------------------------
\end{abstract} 
%---------------------------------------------------------------------

\maketitle

%%%%%%%%%%%%%%%%%%%%%%%%%%%%%%%%%%%%%%%%%%%%%%%%%%%%%%%%%%%%%%%%%%%%%%%
\section{Introduction}\label{sec:introduction}
%%%%%%%%%%%%%%%%%%%%%%%%%%%%%%%%%%%%%%%%%%%%%%%%%%%%%%%%%%%%%%%%%%%%%%%

Fragmentation functions (FFs) encode the nonperturbative transition of
partons to observed hadrons and are essential ingredients in the
factorized description of high-energy hadron-production processes~\cite{Gao:2025hlm,Gao:2024nkz,Gao:2024dbv,AbdulKhalek:2022laj,Borsa:2022vvp,Khalek:2021gxf,Moffat:2021dji}.  
Their universality allows us to combine information from different reactions 
in global QCD analyses, most notably single-inclusive
electron-positron annihilation (SIA), semi-inclusive deep-inelastic
scattering (SIDIS), and inclusive hadron production in hadronic
collisions~\cite{Collins:1989gx,Collins:1981uw,Albino:2008fy}.  In this
framework, perturbatively calculable short-distance hard coefficients are
convoluted with universal long-distance FFs, making the latter
indispensable for precision QCD phenomenology.  While SIA provides clean
constraints on charge-summed FF combinations, SIDIS multiplicities are
crucial for improving flavor and charge separation through their
convolution with target parton distribution functions (PDFs).  Hadron
production in proton-proton collisions, on the other hand, provides
important sensitivity to the gluon FF, especially through high-energy
charged-particle spectra. 

Unidentified charged-hadron FFs are phenomenologically important because
inclusive charged-particle spectra are measured with high precision in
fixed-target, collider, proton-nucleus, and heavy-ion environments.
They enter theoretical predictions for a wide range of observables,
including charged-hadron production in SIA, SIDIS, proton-proton
collisions, and nuclear collision systems. They are also relevant for
future studies at the Electron Ion Collider~\cite{Proceedings:2026xrb,AbdulKhalek:2021gbh}, 
where SIDIS measurements 
will play a central role in constraining the flavor dependence of
hadronization.  In contrast to identified-hadron FFs, which describe the
production of specific hadron species such as pions, kaons, or protons,
unidentified charged-hadron FFs effectively describe the inclusive sum of
all charged hadrons produced in the fragmentation of a given parton.
This makes them especially useful for collider and heavy-ion
applications but also increases the need for accurate global
determinations and reliable uncertainty estimates.

Several determinations of unidentified charged-hadron FFs have appeared
in the literature.  The DSS analysis provided an early global next-to-leading order (NLO)
extraction based on SIA, SIDIS, and proton-proton data~\cite{deFlorian:2007ekg}.  
Recent next-to-next-to-leading order (NNLO) and finite-mass studies of
light charged-hadron fragmentation have also shown the importance of
higher-order corrections and hadron-mass effects in precision FF
determinations~\cite{Soleymaninia:2018uiv,Soleymaninia:2025cvi,Soleymaninia:2020bsq,Gao:2025bko}.
The NNFF1.1h analysis determined unidentified charged-hadron FFs at NLO
using SIA data supplemented by Tevatron and LHC charged-hadron spectra
through Bayesian reweighting, demonstrating in particular the strong
constraining power of hadron-collider data on the gluon
FF~\cite{Bertone:2018ecm}.   The
JAM20-SIDIS analysis performed a simultaneous Monte Carlo extraction of
PDFs and FFs using SIA and SIDIS data for pions, kaons, and
unidentified charged hadrons, emphasizing the role of SIDIS data in
testing universality and constraining flavor dependence~\cite{Moffat:2021dji}.  
More recently, high-energy collider data have
been used to update charged-hadron FFs for modern LHC phenomenology~\cite{Borsa:2023zxk}.  
In addition, our previous SHK22.h analysis
included COMPASS SIDIS multiplicities in a neural-network determination
of unidentified charged-hadron FFs~\cite{Soleymaninia:2022alt}.  These
developments show that unidentified charged-hadron FFs are active
phenomenological inputs whose extraction depends sensitively on the
available experimental information.

The central motivation for the present work is the substantial revision
of the COMPASS SIDIS input relevant for unidentified charged-hadron FFs.
The COMPASS Collaboration has recently published proton-target
multiplicities for charged pions, kaons, and unidentified charged
hadrons, and has subsequently provided revised isoscalar-target
multiplicities in a dedicated addendum~\cite{COMPASS:2024gje,COMPASS:2025bfn}.
The latter supersede the original COMPASS 2017
multiplicities that entered earlier unidentified charged-hadron FF
determinations~\cite{COMPASS:2016xvm,COMPASS:2016crr}. This motivates a
new SIA+SIDIS extraction in which the older COMPASS isoscalar input is
replaced by the modern COMPASS dataset composed of the proton-target
measurement and the revised multiplicities. The
proton-target data provide additional charge-separated flavor
sensitivity, while the revised $^6$LiD-target data provide a more
balanced light-flavor constraint.

The aim of the present work is therefore to revisit unidentified
charged-hadron FFs using the modern COMPASS SIDIS input. We perform a
new global analysis of SIA data together with the COMPASS 2025
proton-target measurement~\cite{COMPASS:2024gje} and the
$^6$LiD-target multiplicities provided in the 2026 COMPASS addendum  for $h^+$ and
$h^-$~\cite{COMPASS:2025bfn}. The analysis is carried out at both NLO and NNLO using the
publicly available \textsc{MontBlanc} framework~\cite{AbdulKhalek:2022laj,MontBlancCode,MontBlancZenodo}.  

At NNLO, the SIA coefficient functions and timelike Dokshitzer--Gribov--Lipatov--Altarelli--Parisi (DGLAP) 
evolution equations~\cite{Altarelli:1977zs,Dokshitzer:1977sg,Gribov:1972ri,Lipatov:1974qm} are 
included at full NNLO accuracy, while the SIDIS coefficient-function
contributions are treated at NNLO within the theoretical framework used
in this analysis. Very recently, exact NNLO QCD coefficient functions for
SIDIS have become available~\cite{Goyal:2024emo,Bonino:2024qbh,Goyal:2023zdi,Bonino:2025qta}, 
making it possible to incorporate
semi-inclusive DIS multiplicities in global FF determinations beyond the
approximate threshold-based treatment used in earlier
studies~\cite{Abele:2021nyo}.

We emphasize that the novelty of the present NNLO analysis should be
understood in the context of the updated COMPASS SIDIS input. Previous
NNLO studies of unidentified charged-hadron FFs have been performed in
the SIA-only context~\cite{Soleymaninia:2018uiv}. Thus, the new element
of the present work is not the use of NNLO corrections alone, but the
incorporation of the COMPASS 2025 proton-target measurement and the
revised multiplicities from the COMPASS 2026
in a global SIA+SIDIS determination at NLO and NNLO. The comparison of NLO and
NNLO FFs is used to assess the perturbative stability of the extracted
FFs within the present theoretical framework.

This paper is organized as follows:  In
Sec.~\ref{sec:Theoretical} we discuss the theoretical framework,
including the SIA and SIDIS factorized expressions, timelike evolution,
and the neural-network parametrization used for the FFs.  In
Sec.~\ref{sec:data} we describe the experimental datasets, the updated
COMPASS input, the uncertainty treatment, and the kinematic selections.
In Sec.~\ref{sec:Qscan}, we study the dependence of the fit quality on the
lower SIDIS scale cut and use this scan to motivate the default choice adopted
in the final fit. 
The main findings and the numerical results are
discussed in Sec.~\ref{sec:results}, including the fit quality, the
description of the COMPASS multiplicities, the comparison between the
NLO and NNLO extractions, and the NLO comparison with NNFF1.1h.  
Finally, we summarize our conclusions and discuss future directions in
Sec.~\ref{sec:Summary}.

%%%%%%%%%%%%%%%%%%%%%%%%%%%%%%%%%%%%%%%%%%%%%%%%%%%%%%%%%%%%%%%%%%%%%%%
\section{Theoretical framework}\label{sec:Theoretical}
%%%%%%%%%%%%%%%%%%%%%%%%%%%%%%%%%%%%%%%%%%%%%%%%%%%%%%%%%%%%%%%%%%%%%%%

In this section, we summarize the theoretical framework used for the
extraction of unidentified charged-hadron FFs
from SIA and SIDIS observables. The analysis
is performed within perturbative QCD using
collinear factorization, in which short-distance partonic cross sections
are calculable in perturbation theory, while the long-distance
hadronization process is encoded in universal FFs~\cite{Collins:1989gx,Collins:1981uw,Albino:2008fy}.
We first introduce the factorized expressions for SIA and SIDIS and then
define the corresponding kinematic variables and the multiplicities used
for the COMPASS data. We also discuss the timelike DGLAP evolution equations,
the perturbative accuracy, the flavor basis, charge-conjugation relations,
and the neural-network parametrization used at the input scale.

%%%%%%%%%%%%%%%%%%%%%%%%%%%%%%%%%%%%%%%%%%%%%%%%%%%%%%%%%%%%%%%%%%%%%%%
\subsection{Collinear factorization}
%%%%%%%%%%%%%%%%%%%%%%%%%%%%%%%%%%%%%%%%%%%%%%%%%%%%%%%%%%%%%%%%%%%%%%%

For the processes considered in this work, the factorized cross sections
may be written schematically as 
%--------------------------------------------------------------------- 
\begin{align}
\sigma_{\rm SIA} &= \hat{\sigma}\otimes D^h,\\
\sigma_{\rm SIDIS} &= \hat{\sigma}\otimes f^N\otimes D^h\,.
\end{align}
%--------------------------------------------------------------------- 
Here, $\hat{\sigma}$ denotes the perturbatively calculable hard-scattering
coefficient, $f^N$ is the parton distribution function (PDF) of the target
nucleon or nucleus $N$, and $D^h$ is the FF into the observed hadron $h$.
More explicitly, $D_i^h$ denotes the FF for a parton of flavor $i$ into the
observed hadron $h$. The symbol $\otimes$ denotes the appropriate
convolution over longitudinal momentum fractions. The universality of FFs
allows the same nonperturbative functions to be constrained simultaneously
using SIA and SIDIS measurements.

Throughout the analysis, $\mu_R$ denotes the renormalization scale, whereas
$\mu_{F}^\mathrm{ini}$ and $\mu_{F}^\mathrm{fin}$ denote the initial-state
and final-state factorization scales, respectively. For compactness, these
scales are identified with a common scale $\mu$ in the expressions below,
%---------------------------------------------------------------------
\begin{equation}
\mu_R = \mu_{F}^{\mathrm{ini}} = \mu_{F}^\mathrm{fin} \equiv \mu,
\end{equation}
%---------------------------------------------------------------------
unless otherwise stated. In SIA, only the final-state factorization scale
is relevant because no PDFs enter the observable.

%%%%%%%%%%%%%%%%%%%%%%%%%%%%%%%%%%%%%%%%%%%%%%%%%%%%%%%%%%%%%%%%%%%%%%%
\subsection{Single-inclusive annihilation}
%%%%%%%%%%%%%%%%%%%%%%%%%%%%%%%%%%%%%%%%%%%%%%%%%%%%%%%%%%%%%%%%%%%%%%%

The SIA process considered in this analysis is 
%---------------------------------------------------------------------
\begin{equation}
e^+e^- \to h^\pm + X ,
\end{equation}
%---------------------------------------------------------------------
where $h^\pm$ denotes the observed positively or negatively charged hadron
and $X$ represents all unobserved particles in the final state. The
annihilation proceeds through an intermediate virtual photon or $Z$ boson
carrying total four-momentum $q$.   
The characteristic hard scale is timelike
and is defined by
%---------------------------------------------------------------------
\begin{equation}
q^2=s\equiv Q^2>0,
\end{equation}
%---------------------------------------------------------------------
where $\sqrt{s}$ is the electron--positron center-of-mass energy.
In the massless-hadron approximation, the SIA scaling variable is defined as
%---------------------------------------------------------------------
\begin{equation}
z = \frac{2p_h\cdot q}{Q^2},
\end{equation}
%---------------------------------------------------------------------
where $p_h$ is the four-momentum of the observed hadron. In the
electron--positron center-of-mass frame, this definition reduces to
$z=2E_h/Q$, where $E_h$ is the hadron energy.

The normalized SIA cross section can be expressed as a convolution of
perturbative coefficient functions and FFs,
%---------------------------------------------------------------------
\begin{equation}
\frac{1}{\sigma_{\rm tot}}\frac{d\sigma^h}{dz}
=
\sum_i
C_i(z,\alpha_s,Q^2,\mu^2)
\otimes
D_i^h(z,\mu^2),
\end{equation}
%---------------------------------------------------------------------
where $\sigma_{\rm tot}$ is the total inclusive hadronic SIA cross section,
$\alpha_s$ is the QCD strong coupling, and the index
$i\in\{q,\bar q,g\}$ runs over the active quark, antiquark, and gluon
flavors. The quantity $C_i$ is the perturbatively calculable coefficient
function associated with the production and subsequent fragmentation of
parton $i$. Its dependence on the renormalization scale is implicit through
$\alpha_s=\alpha_s(\mu_R^2)$. In the notation above, the renormalization and
final-state factorization scales are set equal to the common scale $\mu$.

SIA data provide particularly clean constraints on FFs because no
initial-state PDFs enter the observable. However, SIA measurements alone
have limited power to distinguish quark from antiquark fragmentation. They
constrain the gluon FF mainly through the scale dependence of the cross
sections and through gluon-initiated contributions to the higher-order
coefficient functions%
~\cite{Bertone:2017tyb,Bertone:2018ecm,Soleymaninia:2022alt}.

%%%%%%%%%%%%%%%%%%%%%%%%%%%%%%%%%%%%%%%%%%%%%%%%%%%%%%%%%%%%%%%%%%%%%%%
\subsection{Semi-inclusive DIS multiplicities}
%%%%%%%%%%%%%%%%%%%%%%%%%%%%%%%%%%%%%%%%%%%%%%%%%%%%%%%%%%%%%%%%%%%%%%%

The SIDIS process relevant for the COMPASS measurements is
%---------------------------------------------------------------------
\begin{equation}
\ell(k)+N(p)\to \ell(k')+h^\pm(p_h)+X ,
\end{equation}
%---------------------------------------------------------------------
where $\ell$ denotes the incoming and outgoing charged lepton, $N$ is the
target nucleon or nucleus, and $k$, $k'$, $p$, and $p_h$ are the
four-momenta of the incoming lepton, outgoing lepton, target, and observed
hadron, respectively. The quantity $X$ denotes the unobserved final-state
particles. The four-momentum transferred from the lepton to the target is
%---------------------------------------------------------------------
\begin{equation}
q=k-k'.
\end{equation}
%---------------------------------------------------------------------
In contrast to SIA, the hard scale in SIDIS is spacelike and 
is given by the positive photon virtuality
%---------------------------------------------------------------------
\begin{equation}
Q^2\equiv -q^2>0.
\end{equation}
%---------------------------------------------------------------------

We use the standard dimensionless SIDIS variables
%---------------------------------------------------------------------
\begin{equation}
x = \frac{Q^2}{2p\cdot q}, \qquad
z_h = \frac{p_h\cdot p}{q\cdot p}, \qquad
y = \frac{Q^2}{xs},
\end{equation}
%---------------------------------------------------------------------
where $x$ is the Bjorken scaling variable, $z_h$ is the fraction of the
virtual-photon energy carried by the observed hadron in the target rest
frame, and $y$ is the lepton inelasticity. The variable
%---------------------------------------------------------------------
\begin{equation}
s=(k+p)^2
\end{equation}
%---------------------------------------------------------------------
is the squared lepton--target center-of-mass energy.

The differential SIDIS cross section for charged-hadron production may be
written as
%---------------------------------------------------------------------
\begin{align}
\frac{d\sigma^h}{dx\,dy\,dz_h}
&=
\frac{2\pi\alpha^2}{Q^2}
\Bigg[
\frac{1+(1-y)^2}{y}\,
2F_1^h(x,z_h,Q^2)
\nonumber\\
&\hspace{1.1cm}
+
\frac{2(1-y)}{y}\,
F_L^h(x,z_h,Q^2)
\Bigg],
\end{align}
%---------------------------------------------------------------------
where $\alpha$ is the electromagnetic fine-structure constant, while
$F_1^h$ and $F_L^h$ are the transverse and longitudinal SIDIS structure
functions, respectively.

At NLO, these structure functions can be written in
terms of PDFs, FFs, and perturbative coefficient functions%
~\cite{Altarelli:1979kv,Graudenz:1994dq}:
%---------------------------------------------------------------------
\begin{align}
F_1^h
&=
\frac{1}{2}
\sum_{q,\bar q} e_q^2
\Bigg\{
f_q^N D_q^h
+
\frac{\alpha_s}{2\pi}
\Big[
f_q^N\!\otimes C_{qq}^{1}\!\otimes D_q^h
\nonumber\\
&\hspace{2.0cm}
+
f_q^N\!\otimes C_{gq}^{1}\!\otimes D_g^h
+
f_g^N\!\otimes C_{qg}^{1}\!\otimes D_q^h
\Big]
\Bigg\},
\\[0.2cm]
F_L^h
&=
\frac{\alpha_s}{2\pi}
\sum_{q,\bar q} e_q^2
\Big[
f_q^N\!\otimes C_{qq}^{L}\!\otimes D_q^h
+
f_q^N\!\otimes C_{gq}^{L}\!\otimes D_g^h
\nonumber\\
&\hspace{2.0cm}
+
f_g^N\!\otimes C_{qg}^{L}\!\otimes D_q^h
\Big].
\end{align}
%---------------------------------------------------------------------
Here, $e_q$ is the electric charge of quark or antiquark $q$ in units of
the positron charge, $f_q^N$ and $f_g^N$ are the quark or antiquark and
gluon PDFs of the target, and $D_q^h$ and $D_g^h$ are the corresponding
quark or antiquark and gluon FFs. The sum $\sum_{q,\bar q}$ runs over all
active quark and antiquark flavors.

The coefficient functions $C_{ab}^{1}$ and $C_{ab}^{L}$ correspond to the
structure functions $F_1^h$ and $F_L^h$, respectively. In the index
convention used above, the first index labels the fragmenting final-state
parton and the second index labels the incoming parton. Thus, for example,
$C_{gq}$ describes a channel with an incoming quark and a fragmenting
final-state gluon, whereas $C_{qg}$ corresponds to an incoming gluon and a
fragmenting quark. The dependence of the PDFs, FFs, coefficient functions,
and strong coupling on the factorization and renormalization scales is
suppressed in these equations for readability.

The double convolution appearing in the structure functions is defined as
%---------------------------------------------------------------------
\begin{align}
\big[f\otimes C\otimes D\big](x,z_h)
&\equiv
\int_x^1\frac{d\xi}{\xi}
\int_{z_h}^1\frac{d\zeta}{\zeta}
\nonumber\\
&\quad\times
f\!\left(\frac{x}{\xi}\right)
C(\xi,\zeta)
D\!\left(\frac{z_h}{\zeta}\right),
\end{align}
%---------------------------------------------------------------------
where $\xi$ and $\zeta$ are the longitudinal momentum fractions entering
the initial-state and final-state convolutions, respectively.

The COMPASS multiplicity is defined as the ratio of the semi-inclusive
hadron-production cross section to the corresponding inclusive
deep-inelastic-scattering (DIS) cross section,
%---------------------------------------------------------------------
\begin{equation}
\frac{dM^h(x,z_h,Q^2)}{dz_h}
=
\frac{
d^3\sigma^h(x,z_h,Q^2)/dx\,dQ^2\,dz_h
}{
d^2\sigma^{\rm DIS}(x,Q^2)/dx\,dQ^2
}.
\end{equation}
%---------------------------------------------------------------------
At leading order (LO), this expression reduces to
%---------------------------------------------------------------------
\begin{equation}
\frac{dM^h}{dz_h}
=
\frac{
\displaystyle
\sum_{q,\bar q}
e_q^2
f_q^N(x,Q^2)
D_q^h(z_h,Q^2)
}{
\displaystyle
\sum_{q,\bar q}
e_q^2
f_q^N(x,Q^2)
}.
\end{equation}
%---------------------------------------------------------------------
The different electric-charge and PDF weights associated with the quark
and antiquark flavors make the charge-separated SIDIS multiplicities for
$h^+$ and $h^-$ particularly important for determining the flavor
dependence of the FFs.

%%%%%%%%%%%%%%%%%%%%%%%%%%%%%%%%%%%%%%%%%%%%%%%%%%%%%%%%%%%%%%%%%%%%%%%
\subsection{Timelike DGLAP evolution and perturbative order}
%%%%%%%%%%%%%%%%%%%%%%%%%%%%%%%%%%%%%%%%%%%%%%%%%%%%%%%%%%%%%%%%%%%%%%%

The dependence of the FFs on the factorization scale is governed by the
timelike DGLAP evolution equations,
%---------------------------------------------------------------------
\begin{equation}
\frac{dD_i^h(z,\mu^2)}{d\ln\mu^2}
=
\frac{\alpha_s(\mu^2)}{2\pi}
\sum_j
\int_z^1 \frac{d\xi}{\xi}
P^T_{ji}(\xi,\alpha_s)
D_j^h\!\left(\frac{z}{\xi},\mu^2\right).
\end{equation}
%---------------------------------------------------------------------
Here, $i$ and $j$ label quarks, antiquarks, and the gluon, and $P^T_{ji}$
is the timelike splitting function that couples the FF $D_j^h$ to the
scale evolution of $D_i^h$. The superscript $T$ distinguishes the timelike
evolution kernels relevant for fragmentation from the spacelike kernels
entering PDF evolution.

The timelike splitting functions are expanded perturbatively as
%---------------------------------------------------------------------
\begin{equation}
P^T_{ji}(z,\alpha_s)
=
P^{T,(0)}_{ji}(z)
+
\frac{\alpha_s}{2\pi}P^{T,(1)}_{ji}(z)
+
\left(\frac{\alpha_s}{2\pi}\right)^2
P^{T,(2)}_{ji}(z)
+\mathcal{O}(\alpha_s^3).
\end{equation}
%---------------------------------------------------------------------
The leading-order evolution involves only $P^{T,(0)}$, while the NLO
and NNLO evolutions additionally include $P^{T,(1)}$ and $P^{T,(2)}$,
respectively.

The NNLO timelike evolution used in modern FF extractions is based on the
known NNLO timelike splitting functions and their implementations in
evolution libraries%
~\cite{Mitov:2006ic,Moch:2007tx,Almasy:2011eq,Bertone:2015cwa}.
For SIA, the perturbative coefficient functions are available through
NNLO%
~\cite{Rijken:1996ns,Rijken:1996vr,Rijken:1996npa}.

Full fixed-order NNLO QCD coefficient functions for unpolarized SIDIS have
also become available%
~\cite{Goyal:2024emo,Bonino:2024qbh,Goyal:2023zdi,Bonino:2025qta}.
In the present analysis, NNLO SIDIS coefficient functions are
incorporated through the implementation used in \textsc{MontBlanc}~\cite{Abele:2021nyo,AbdulKhalek:2022laj,MontBlancCode}.
Consequently, the NNLO analysis consistently combines NNLO timelike
evolution with coefficient functions through $\mathcal{O}(\alpha_s^2)$
for both SIA and SIDIS observables. The explicit expressions above display
the NLO partonic decomposition for clarity; at NNLO, the additional
two-loop coefficient-function contributions and partonic channels are
included numerically in the theoretical predictions.

%%%%%%%%%%%%%%%%%%%%%%%%%%%%%%%%%%%%%%%%%%%%%%%%%%%%%%%%%%%%%%%%%%%%%%%
\subsection{Flavor basis and charge conjugation}
%%%%%%%%%%%%%%%%%%%%%%%%%%%%%%%%%%%%%%%%%%%%%%%%%%%%%%%%%%%%%%%%%%%%%%%

For positively charged unidentified hadrons, we parametrize the
independent FF combinations at the input scale $Q_0=5~\mathrm{GeV}$ as 
%---------------------------------------------------------------------
\begin{equation} 
D_u^{h^+},\quad
D_{\bar u}^{h^+},\quad
D_{d+s}^{h^+},\quad
D_{\bar d + \bar s}^{h^+},\quad
D_{c^+}^{h^+},\quad
D_{b^+}^{h^+},\quad
D_g^{h^+}.
\end{equation} 
%---------------------------------------------------------------------
The strange-quark contribution is included in the SIDIS and SIA flavor
sums, but it is not parametrized as an additional independent
neural-network output in the present fit. Instead, the strange and
antistrange FFs are tied to the light-antiquark sector through the same
flavor-basis assumption used in the fit implementation and in the
previous SHK22.h analysis~\cite{Soleymaninia:2022alt}.  This reduces the number of independent
nonperturbative degrees of freedom in a sector that is only weakly
constrained by the present unidentified charged-hadron data. 

The FFs for negatively charged hadrons are obtained using charge
conjugation, 
%---------------------------------------------------------------------
\begin{equation}
D_q^{h^-}(z,Q)=D_{\bar q}^{h^+}(z,Q),\qquad
D_{\bar q}^{h^-}(z,Q)=D_q^{h^+}(z,Q),
\end{equation}
%---------------------------------------------------------------------
and
%---------------------------------------------------------------------
\begin{equation}
D_g^{h^-}(z,Q)=D_g^{h^+}(z,Q).
\end{equation}
%---------------------------------------------------------------------
For the initial-state parton distributions entering the SIDIS
predictions, we use the NNPDF4.0 sets~\cite{NNPDF:2021njg},
namely \texttt{NNPDF40\_nlo\_pch\_as\_01180} and
\texttt{NNPDF40\_nnlo\_pch\_as\_01180} for the NLO and NNLO
analyses, respectively. The strong coupling is fixed to
$\alpha_s(M_Z)=0.118$, with $M_Z=91.1876~\mathrm{GeV}$,
consistently with the corresponding PDF ensembles. 
The heavy-quark thresholds are set to
$Q_c=1.51~\mathrm{GeV}$ and $Q_b=4.92~\mathrm{GeV}$ for charm and
bottom, respectively.

%%%%%%%%%%%%%%%%%%%%%%%%%%%%%%%%%%%%%%%%%%%%%%%%%%%%%%%%%%%%%%%%%%%%%%%
\subsection{Parametrization and uncertainty propagation}
%%%%%%%%%%%%%%%%%%%%%%%%%%%%%%%%%%%%%%%%%%%%%%%%%%%%%%%%%%%%%%%%%%%%%%%

The FFs are parametrized at $Q_0=5~\mathrm{GeV}$ using a neural-network
representation, following previous neural-network FF extractions and the
\textsc{MontBlanc} framework~\cite{Bertone:2017tyb,Soleymaninia:2022alt,AbdulKhalek:2022laj,MontBlancCode, MontBlancZenodo}. 
For each independent flavor combination, we write
%---------------------------------------------------------------------
\begin{equation}
zD_i^{h^+}(z,Q_0)=
\left[
N_i(z,\theta)-N_i(1,\theta)
\right]^2 ,
\end{equation}
%---------------------------------------------------------------------
where $N_i(z,\theta)$ is the neural-network output and $\theta$ denotes
the set of internal parameters. The subtraction enforces the endpoint
condition at $z=1$, while the squared form guarantees positivity at the
input scale.

The baseline architecture contains one hidden layer with 20 neurons.
Experimental uncertainties are propagated using Monte Carlo replicas. In
the present analysis we use $N_{\rm rep}=200$ replicas. In the
\textsc{MontBlanc} implementation, minimization uses the \textsc{Ceres
Solver} package and analytic neural-network derivatives from
NNAD~\cite{Agarwal:Ceres,AbdulKhalek:2020uza,AbdulKhalek:2022laj}.

%%%%%%%%%%%%%%%%%%%%%%%%%%%%%%%%%%%%%%%%%%%%%%%%%%%%%%%%%%%%%%%%%%%%%%%
\section{Experimental datasets and kinematic cuts}\label{sec:data}
%%%%%%%%%%%%%%%%%%%%%%%%%%%%%%%%%%%%%%%%%%%%%%%%%%%%%%%%%%%%%%%%%%%%%%%

The present analysis uses SIA data and COMPASS SIDIS multiplicities for
unidentified charged hadrons. To isolate the impact of the new and
revised COMPASS measurements, we use the same SIA baseline as in the
previous SHK22.h analysis~\cite{Soleymaninia:2022alt} and update the COMPASS SIDIS input~\cite{COMPASS:2024gje,COMPASS:2025bfn}.

\begin{figure*}[t]  
\centering
\includegraphics[width=0.70\textwidth]{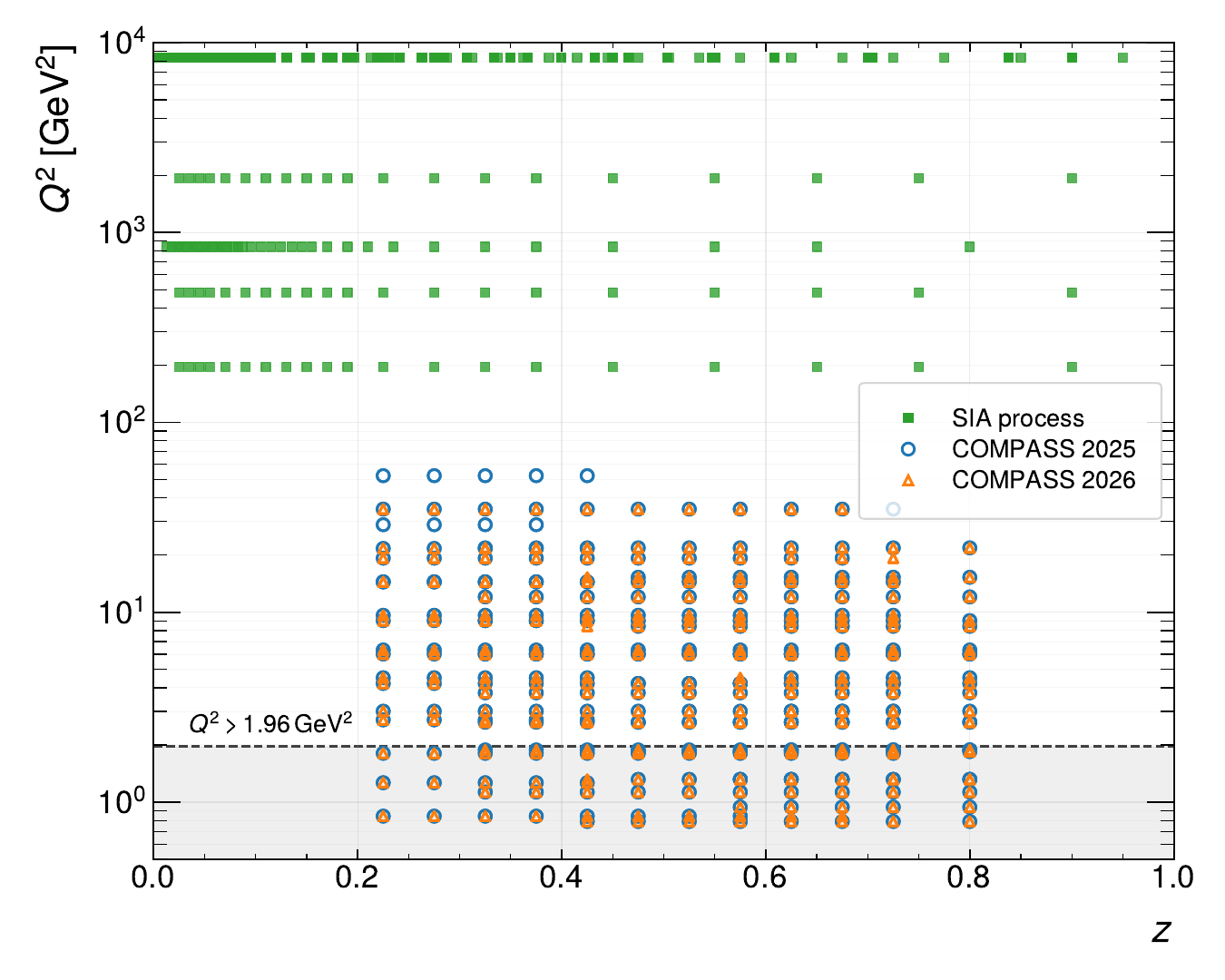}
\caption{\small
Kinematic coverage in the \((z,Q^2)\) plane for the datasets included in
the present analysis. The COMPASS 2025 proton-target multiplicities and
the revised isoscalar-target multiplicities from the 2026 COMPASS
addendum for unidentified charged hadrons are shown together with the SIA
datasets. The dashed horizontal line indicates the baseline perturbative
cut, \(Q^2>1.96~\mathrm{GeV}^2\), corresponding to
\(Q>1.4~\mathrm{GeV}\).
}
\label{fig:q2-z-coverage}
\end{figure*}

The corresponding kinematic coverage in the $(z,Q^{2})$ plane is shown in
Fig.~\ref{fig:q2-z-coverage}. 
The dashed horizontal line indicates the baseline cut $Q^{2}>1.96~\mathrm{GeV}^{2}$, corresponding to
$Q>1.4~\mathrm{GeV}$.

%---------------------------------------------------------------------
%---------------------------------------------------------------------

%%%%%%%%%%%%%%%%%%%%%%%%%%%%%%%%%%%%%%%%%%%%%%%%%%%%%%%%%%%%%%%%%%%%%%%
\subsection{SIA data sets}
%%%%%%%%%%%%%%%%%%%%%%%%%%%%%%%%%%%%%%%%%%%%%%%%%%%%%%%%%%%%%%%%%%%%%%%

The SIA dataset consists of inclusive and flavor-tagged measurements of
unidentified charged-hadron production in
%---------------------------------------------------------------------
\begin{equation}
e^+e^- \to h^\pm + X
\end{equation}
%---------------------------------------------------------------------
from the TASSO, TPC, ALEPH, DELPHI, OPAL, and SLD
Collaborations~\cite{TASSO:1980dyh,TASSO:1983cre,TASSO:1989orr,TPCTwoGamma:1988yjh,ALEPH:1994cbg,DELPHI:1998cgx,OPAL:1994zan,SLD:2003ogn}. 
These measurements provide the cleanest constraints on charge-summed
fragmentation combinations, since no PDFs enter
the SIA observable.  The flavor-tagged data taken at the $Z$ pole are
particularly important for separating light-, charm-, and bottom-quark
fragmentation components. 

For measurements taken at the $Z$ pole we retain
$0.02\le z\le0.9$, while for lower-energy SIA data we impose
$0.075\le z\le0.9$~\cite{Soleymaninia:2022alt,AbdulKhalek:2022laj}. 
The more restrictive lower cut applied to the lower-energy data reduces
the impact of the small-$z$ region, where fixed-order calculations are
more sensitive to small-$z$ logarithms, hadron-mass effects, and other
effects beyond the leading-power massless approximation.  The upper cut
removes the endpoint region, where threshold logarithms and experimental
limitations become increasingly important.

%%%%%%%%%%%%%%%%%%%%%%%%%%%%%%%%%%%%%%%%%%%%%%%%%%%%%%%%%%%%%%%%%%%%%%%
\subsection{COMPASS SIDIS data}  
%%%%%%%%%%%%%%%%%%%%%%%%%%%%%%%%%%%%%%%%%%%%%%%%%%%%%%%%%%%%%%%%%%%%%%%

The central new ingredient is the use of the COMPASS 2025
proton-target measurement~\cite{COMPASS:2024gje} together with the revised
$^{6}\mathrm{LiD}$-target multiplicities provided in the 2026
COMPASS ~\cite{COMPASS:2025bfn}.
The new proton-target measurement provides charge-separated multiplicities for
$\pi^\pm$, $K^\pm$, and $h^\pm$ from a liquid hydrogen target, covering
%---------------------------------------------------------------------
\begin{equation}
\begin{aligned}
Q^2 &> 1~\mathrm{GeV}^2, &
0.004 &< x < 0.4, \\
0.1 &< y < 0.7, &
0.2 &< z < 0.85 .
\end{aligned}
\end{equation}
%---------------------------------------------------------------------
The data referred to here as the COMPASS isoscalar-target data were
measured using a solid-state $^{6}\mathrm{LiD}$ target. Owing to small
admixtures of H, $^{3}\mathrm{He}$, and $^{7}\mathrm{Li}$, the target
has an effective neutron excess of approximately $0.2\%$ and is
therefore only approximately isoscalar~\cite{COMPASS:2025bfn}. Following the
standard COMPASS treatment, we approximate the target as effectively
isoscalar in the theoretical calculation. The 2026 COMPASS 
does not correspond to a new target measurement; rather, it provides
revised multiplicities for the same $^{6}\mathrm{LiD}$ data using
DJANGOH-based  quantum electrodynamics (QED)  radiative corrections. 
In the following, we refer to this dataset as the revised $^{6}\mathrm{LiD}$-target data.

The use of the revised multiplicities is an
important ingredient of the present analysis. In the COMPASS
2026, the earlier TERAD-based treatment of the
radiative corrections was replaced by correction factors derived
with the DJANGOH Monte Carlo generator, following the
strategy used in the recent proton-target measurement. This
improved treatment leads to non-negligible changes in some
kinematic regions; in particular, the COMPASS addendum
reports corrections up to about \(12\%\) larger in the low-\(x\),
high-\(z\) region compared with the previously applied ones.
Using the revised isoscalar multiplicities therefore ensures a
consistent radiative-correction treatment of the proton- and
isoscalar-target COMPASS data sets and avoids combining the
new proton data with superseded isoscalar multiplicities.
           
For unidentified charged hadrons, the pion mass is assumed in the
reconstruction of the hadron energy fraction~\cite{COMPASS:2024gje}. 
The fit uses the published unidentified charged-hadron multiplicities in
the stated SIDIS phase space. The additional particle-identification and
acceptance requirements associated with the COMPASS RICH detector are
part of the experimental treatment of the identified-hadron samples and
of the parent measurement~\cite{COMPASS:2024gje,Abbon:2011zza}; they are
not introduced here as separate cuts beyond the published unidentified
charged-hadron multiplicities used in the fit.

The COMPASS 2026 updates the older $^{6}\mathrm{LiD}$-target
multiplicities for charged pions, charged kaons, and unidentified charged
hadrons~\cite{COMPASS:2025bfn}. The revised multiplicities are
obtained by removing the earlier TERAD-based corrections and applying
DJANGOH-derived correction factors~\cite{Akhundov:1996,Aschenauer:2013woa,COMPASS:2024gje,COMPASS:2025bfn}. 
The updated isoscalar results supersede the previous COMPASS 2017
measurements~\cite{COMPASS:2016xvm,COMPASS:2016crr,COMPASS:2025bfn}. 
In the present analysis, this replacement is
essential for combining the COMPASS proton-target and $^{6}\mathrm{LiD}$-target
measurements on a consistent radiative-correction footing.

For the FF extraction, we use multiplicities corrected for QED radiation
and diffractive vector-meson contributions. Hadrons from diffractive
vector-meson decays do not correspond to the independent parton
fragmentation mechanism described by standard collinear FFs. For
unidentified charged hadrons and pions, the dominant contamination is
associated with $\rho^0$ production, while for kaons it is mainly from
$\phi$ decays~\cite{Sandacz:2012at,COMPASS:2024gje}.

%%%%%%%%%%%%%%%%%%%%%%%%%%%%%%%%%%%%%%%%%%%%%%%%%%%%%%%%%%%%%%%%%%%%%%%
\subsection{Kinematic cuts and uncertainty treatment}
%%%%%%%%%%%%%%%%%%%%%%%%%%%%%%%%%%%%%%%%%%%%%%%%%%%%%%%%%%%%%%%%%%%%%%%

Small-$z$ and low-$Q$ regions are theoretically delicate because of
small-$z$ logarithms, hadron-mass effects, target-mass effects,
higher-twist contributions, and possible limitations of the
current-fragmentation approximation~\cite{Accardi:2014qda,Guerrero:2015wha,Boglione:2016bph,Boglione:2019nwk}. 
For COMPASS SIDIS multiplicities we retain the published range
$0.2<z<0.85$ and impose a lower cut on $Q$, selected by the scan described
in Sec.~\ref{sec:Qscan}.

The COMPASS systematic uncertainties contain a large bin-to-bin correlated
component. Following the COMPASS recommendation and the treatment used in
the MAPFF/\textsc{MontBlanc} framework, we take $80\%$ of the systematic
uncertainty as correlated and add the remaining uncorrelated component in
quadrature with the statistical uncertainty~\cite{COMPASS:2024gje,COMPASS:2025bfn,AbdulKhalek:2022laj}. 
Thus, the uncorrelated uncertainty entering the diagonal part of the
covariance matrix is taken to be
\begin{equation}
\sigma_{\rm unc}
=
\sqrt{\sigma_{\rm stat}^2+(0.6\,\sigma_{\rm syst})^2},  \nonumber
\end{equation}
where the factor \(0.6=\sqrt{1-0.8^2}\) corresponds to the remaining
uncorrelated fraction of the systematic uncertainty. The correlated
systematic component is then
\begin{equation}
\sigma_{\rm corr}=0.8\,\sigma_{\rm syst}.    \nonumber
\end{equation}
%

%%%%%%%%%%%%%%%%%%%%%%%%%%%%%%%%%%%%%%%%%%%%%%%%%%%%%%%%%%%%%%%%%%%%%%%
\section{Selection of the lower SIDIS scale cut} \label{sec:Qscan}
%%%%%%%%%%%%%%%%%%%%%%%%%%%%%%%%%%%%%%%%%%%%%%%%%%%%%%%%%%%%%%%%%%%%%%%

The COMPASS multiplicities extend to relatively low photon virtualities,
where a fixed-order leading-power description may become less reliable.
In this region, higher-twist effects, target-mass and hadron-mass
corrections, and limitations of the current-fragmentation approximation
may become increasingly important. Since the present analysis includes both the 
COMPASS 2025 proton-target
measurement and the revised $^{6}\mathrm{LiD}$-target 
multiplicities provided in the 2026 COMPASS, 
it is important to determine a lower cut on the hard scale that  
provides a stable perturbative description 
without unnecessarily removing SIDIS information from the fit. 

We therefore perform a scan in the lower SIDIS scale cut $Q_{\min}\equiv Q_{\rm cut}$
using the values $Q_{\rm cut}=1.0,\ 1.2,\ 1.4,\ 1.7,\ 1.8,\ 2.0~\mathrm{GeV}$. 
For each value of $Q_{\rm cut}$, the global fit is repeated at both NLO
and NNLO accuracy. The scan is performed separately for the
charge-separated COMPASS $h^+$ and $h^-$ samples from the new
proton-target data and from the revised isoscalar-target data, and also
for the total SIDIS contribution.

\begin{figure*}[t] 
\centering 
\includegraphics[width=0.480\textwidth]{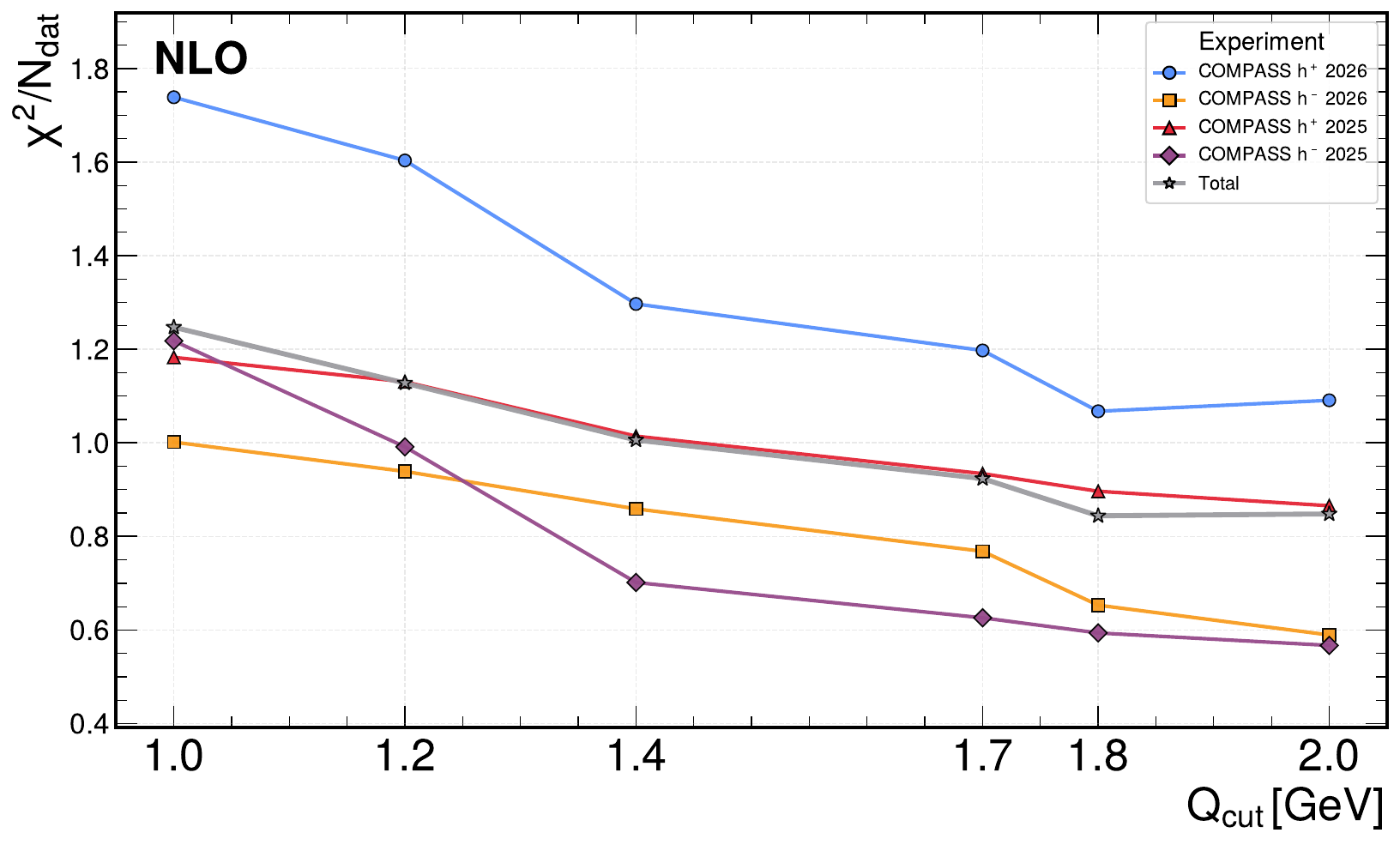}
\includegraphics[width=0.480\textwidth]
{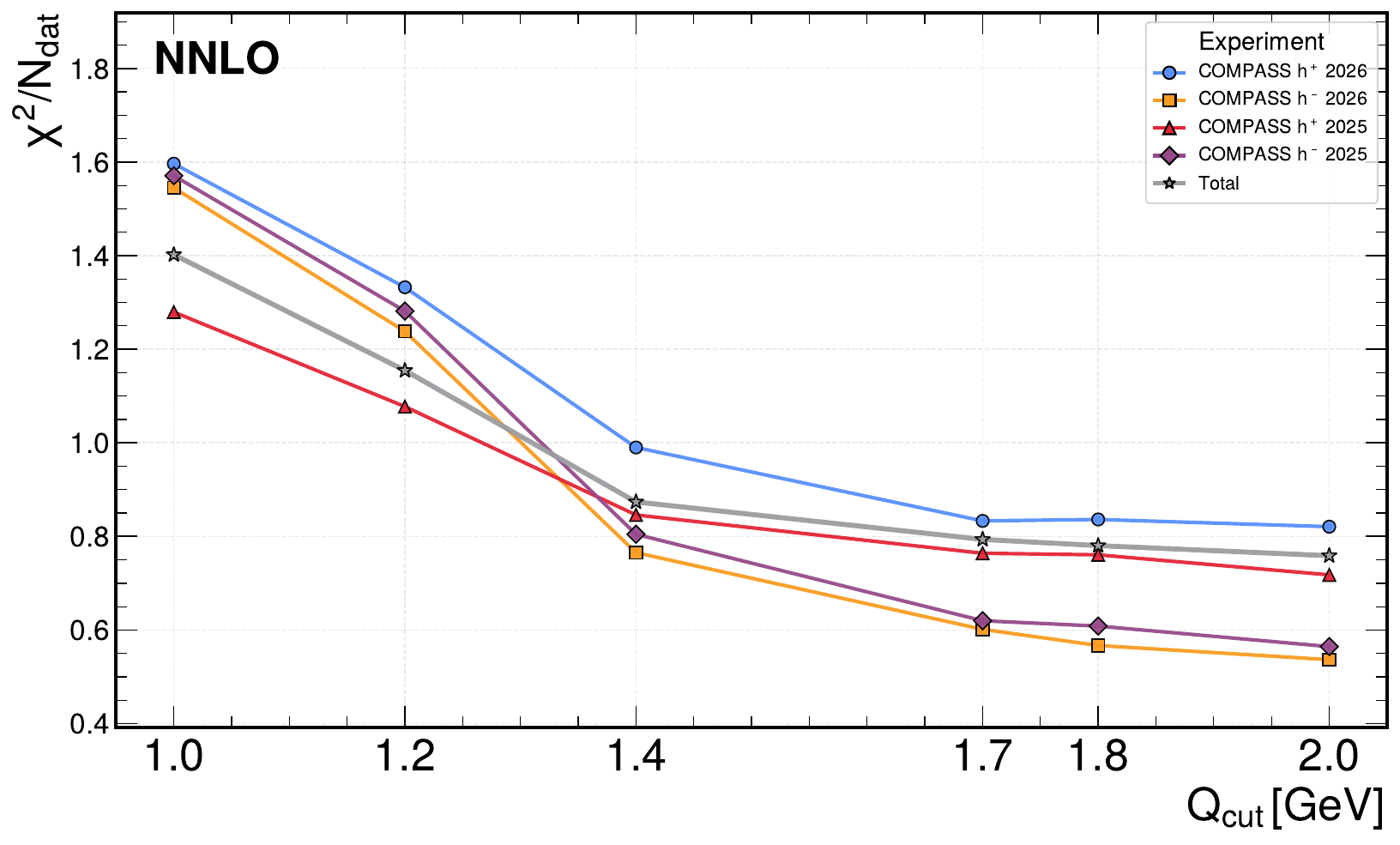}
\caption{\small 
Values of \(\chi^2/N_{\rm dat}\) as a function of the lower SIDIS scale
cut \(Q_{\rm cut}\) for the COMPASS unidentified charged-hadron
multiplicities. 
Results are shown at both NLO (left) and NNLO (right) for the COMPASS
2025 proton-target \(h^+\) and \(h^-\) multiplicities, revised $^{6}\mathrm{LiD}$-target \(h^+\) and \(h^-\) multiplicities from the
COMPASS 2026, and the corresponding total SIDIS contribution. } 
\label{fig:Qcut-chi2-NLO-NNLO} 
\end{figure*}

The results are shown in Fig.~\ref{fig:Qcut-chi2-NLO-NNLO}. At the
lowest cut, $Q_{\rm cut}=1.0~\mathrm{GeV}$, the fit includes the largest
number of COMPASS SIDIS points, but the corresponding values of
$\chi^2/N_{\rm dat}$ are visibly larger for several subsets. This is
particularly evident for the revised COMPASS $h^+$ 2026 data, and it is
also reflected in the total contribution. Increasing the cut from
$1.0$ to $1.2~\mathrm{GeV}$ improves the description, but some remaining
tension is still visible, especially in the low-$Q$ region. A further increase to $Q_{\rm cut}=1.4~\mathrm{GeV}$
leads to a substantial improvement of the total $\chi^2/N_{\rm dat}$ at
both perturbative orders. At this value, the description of the new
COMPASS 2025 data and the revised COMPASS 2026
one becomes stable and acceptable. The improvement is
seen not only in the total contribution, but also in the individual
charge-separated subsets. In particular, the larger tension observed at
lower cuts is significantly reduced, while a large fraction of the SIDIS
data is still retained in the fit.

For larger values of the cut, such as
$Q_{\rm cut}=1.7$, $1.8$, and $2.0~\mathrm{GeV}$, the total
$\chi^2/N_{\rm dat}$ decreases further or remains comparably stable.
However, these more conservative choices remove a larger fraction of the
COMPASS SIDIS data. Since the charge-separated SIDIS multiplicities are
the main source of flavor sensitivity in the present analysis, applying
too high a cut would weaken the constraining power of the fit, especially
for the light-quark and antiquark FFs. We therefore choose
$Q_{\min}=1.4~\mathrm{GeV}$ as the default scale cut for both the NLO and NNLO analyses. 
This choice represents the
best compromise in the present study: it removes the least stable
low-$Q$ region, yields a significant improvement in the fit quality, and
preserves a substantial amount of the new and revised COMPASS SIDIS
information. The same value of $Q_{\min}$ is used for both the NLO and
NNLO fits. 

%---------------------------------------------------------------------
%---------------------------------------------------------------------

%%%%%%%%%%%%%%%%%%%%%%%%%%%%%%%%%%%%%%%%%%%%%%%%%%%%%%%%%%%%%%%%%%%%%%%
\section{Results}\label{sec:results}
%%%%%%%%%%%%%%%%%%%%%%%%%%%%%%%%%%%%%%%%%%%%%%%%%%%%%%%%%%%%%%%%%%%%%%%

In this section we present the main results of the global QCD analysis.
We discuss the quality of the fit, the description of the modern COMPASS
SIDIS multiplicities, and the resulting unidentified charged-hadron FFs. 
Particular attention is given to the impact of the updated COMPASS input
on the light-quark and antiquark fragmentation sectors. We also use the
NLO-NNLO comparison to assess the perturbative stability of the
extraction and compare the present NLO result with NNFF1.1h.

%%%%%%%%%%%%%%%%%%%%%%%%%%%%%%%%%%%%%%%%%%%%%%%%%%%%%%%%%%%%%%%%%%%%%%%
\subsection{Fit quality}
%%%%%%%%%%%%%%%%%%%%%%%%%%%%%%%%%%%%%%%%%%%%%%%%%%%%%%%%%%%%%%%%%%%%%%%

\begin{figure*}[t] 
\centering
\includegraphics[width=0.70\textwidth]{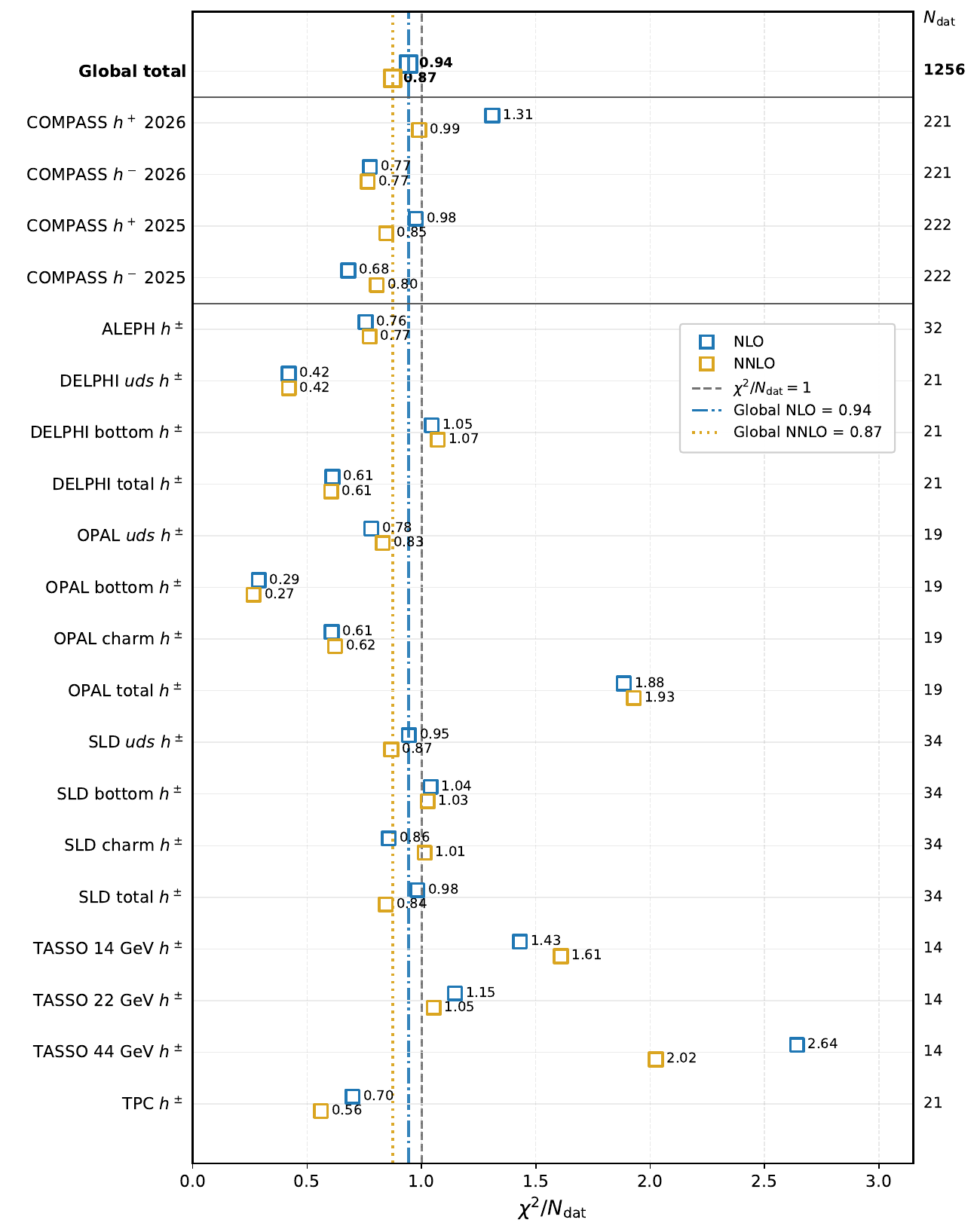}
\caption{\small
Dataset-level fit quality for the final global analysis. The figure shows
$\chi^2/N_{\rm dat}$ for the individual SIA datasets, the COMPASS 2025
proton-target multiplicities, the revised COMPASS 2026
$^{6}\mathrm{LiD}$-target multiplicities, and the global total. The NLO and
NNLO results are displayed using blue and orange open-square markers,
respectively. The corresponding global-fit values are indicated by vertical
blue dashed and orange dotted lines, while the gray dashed line marks the
reference value $\chi^2/N_{\rm dat}=1$. The number of data points,
$N_{\rm dat}$, is shown for each dataset. } 
\label{fig:chi2-circular}
\end{figure*}

The quality of the final fits is summarized in
Fig.~\ref{fig:chi2-circular}, where we show the dataset-level values of  
$\chi^2/N_{\rm dat}$ for the SIA datasets, the COMPASS SIDIS
multiplicities, and the global total. The final dataset contains
%---------------------------------------------------------------------
\begin{equation}
N_{\rm dat}^{\rm SIA}=370,\qquad \nonumber
N_{\rm dat}^{\rm SIDIS}=886,\qquad \nonumber
N_{\rm dat}^{\rm tot}=1256\,. \nonumber
\end{equation}
%---------------------------------------------------------------------
The corresponding global fit qualities are
%---------------------------------------------------------------------
\begin{equation}
\left(\chi^2/N_{\rm dat}\right)_{\rm NLO}=0.943,\nonumber
\qquad
\left(\chi^2/N_{\rm dat}\right)_{\rm NNLO}=0.873.\nonumber
\end{equation}
%---------------------------------------------------------------------
These values show that the combined SIA and modern COMPASS SIDIS
datasets are described well at both perturbative orders. The smaller
global value obtained at NNLO indicates a modest improvement in the
overall fit quality. This improvement should be viewed as an indication
of perturbative stability within the theoretical setup adopted in our QCD analysis. 

Figure~\ref{fig:chi2-circular} provides a compact overview of the
dataset-by-dataset fit quality. For each individual SIDIS and SIA dataset,
the values of $\chi^2/N_{\rm dat}$ are shown at NLO and NNLO using blue and
orange open-square markers, respectively. The vertical blue dashed and orange
dotted lines indicate the corresponding global-fit values, while the gray
dashed line marks the reference value $\chi^2/N_{\rm dat}=1$. The number of
data points, $N_{\rm dat}$, is also indicated for each dataset. The COMPASS
subsets comprise the 2025 proton-target and revised 2026 $^{6}\mathrm{LiD}$-target
multiplicities for both $h^+$ and $h^-$.
The figure shows that the SIDIS datasets are generally
well described and do not dominate the total $\chi^2$. In particular,
the revised COMPASS 2026 $h^+$ contribution decreases noticeably when
going from NLO to NNLO, while the other COMPASS subsets remain at an
acceptable level. 
Nevertheless, the combined SIA contribution remains below unity at both
perturbative orders,
%---------------------------------------------------------------------
\begin{equation}
\left(\chi^2/N_{\rm dat}\right)_{\rm SIA}^{\rm NLO}=0.954, \nonumber
\qquad
\left(\chi^2/N_{\rm dat}\right)_{\rm SIA}^{\rm NNLO}=0.927. \nonumber
\end{equation}
%---------------------------------------------------------------------
Similarly, the total SIDIS contribution decreases from $\left(\chi^2/N_{\rm dat}\right)_{\rm SIDIS}^{\rm NLO}=0.934$
to $\left(\chi^2/N_{\rm dat}\right)_{\rm SIDIS}^{\rm NNLO}=0.851.$ 

The good description of both sectors shows that the modern COMPASS SIDIS
input can be incorporated consistently with the SIA data in a single
global extraction.  The NNLO improvement should be regarded as a
perturbative-stability indication within the present theoretical
framework.

%---------------------------------------------------------------------
%---------------------------------------------------------------------

%%%%%%%%%%%%%%%%%%%%%%%%%%%%%%%%%%%%%%%%%%%%%%%%%%%%%%%%%%%%%%%%%%%%%%%
\subsection{Description of COMPASS multiplicities}\label{subsec:compass-multiplicities}
%%%%%%%%%%%%%%%%%%%%%%%%%%%%%%%%%%%%%%%%%%%%%%%%%%%%%%%%%%%%%%%%%%%%%%%

\begin{figure*}[t]
\centering
\includegraphics[width=0.98\textwidth]{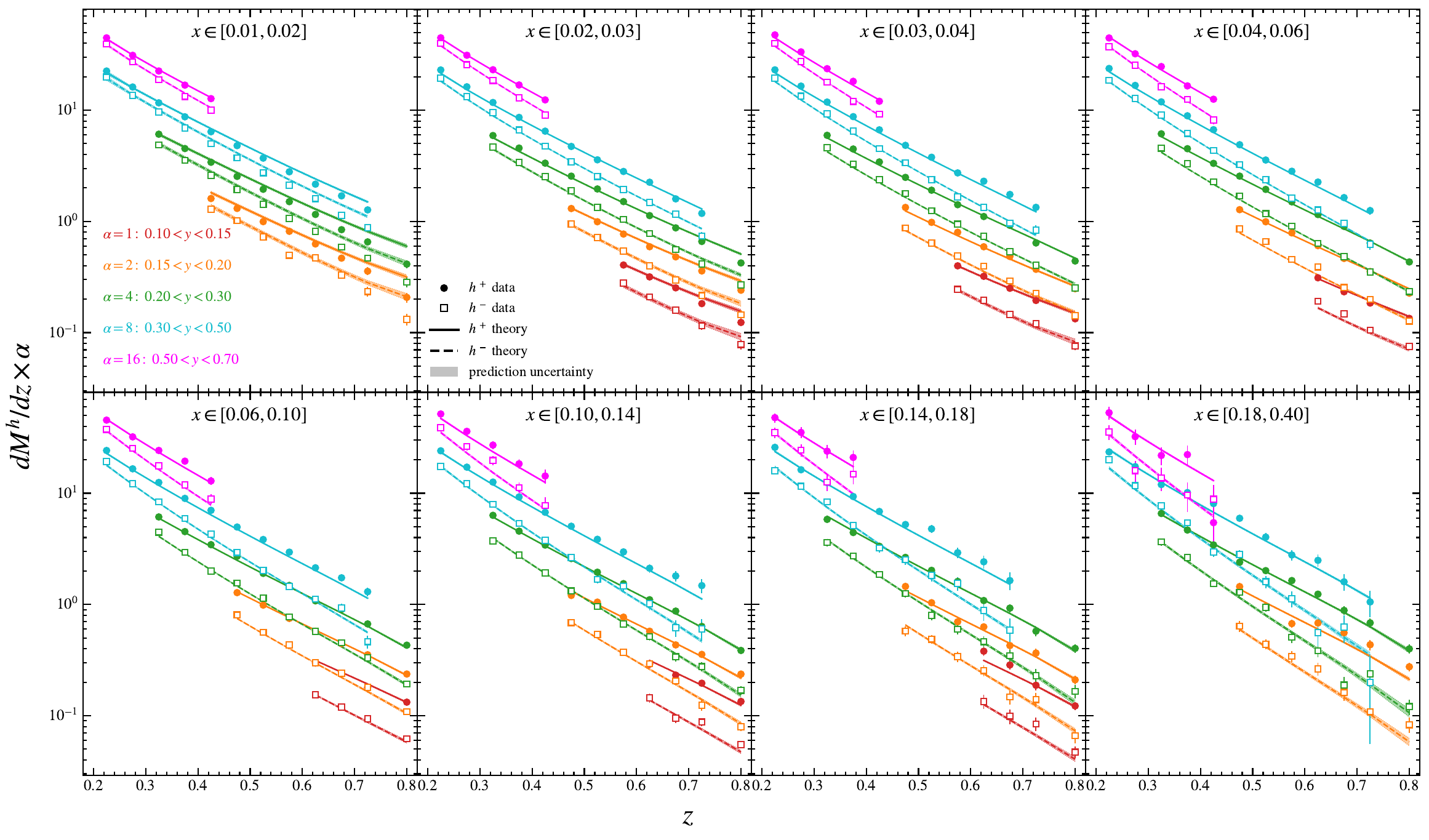}
\caption{\small
COMPASS 2025 proton-target multiplicities for unidentified charged
hadrons~\cite{COMPASS:2024gje} compared with the final fitted theory
predictions. The multiplicities are shown as functions of $z$ in bins of
$x$ and $y$. Filled circles denote $h^+$ data and open squares denote
$h^-$ data. Solid curves show the predictions for $h^+$, while dashed
curves show the corresponding predictions for $h^-$. The vertical
bars represent the experimental uncorrelated uncertainties used in the
fit, and the shaded bands indicate the one-standard-deviation prediction
uncertainty propagated from the fitted FF replicas. The different $y$
intervals are separated by multiplicative factors $\alpha$, as indicated
in the first panel.
}
\label{fig:compass2025-hpm}
\end{figure*}

\begin{figure*}[t]
\centering
\includegraphics[width=0.98\textwidth]{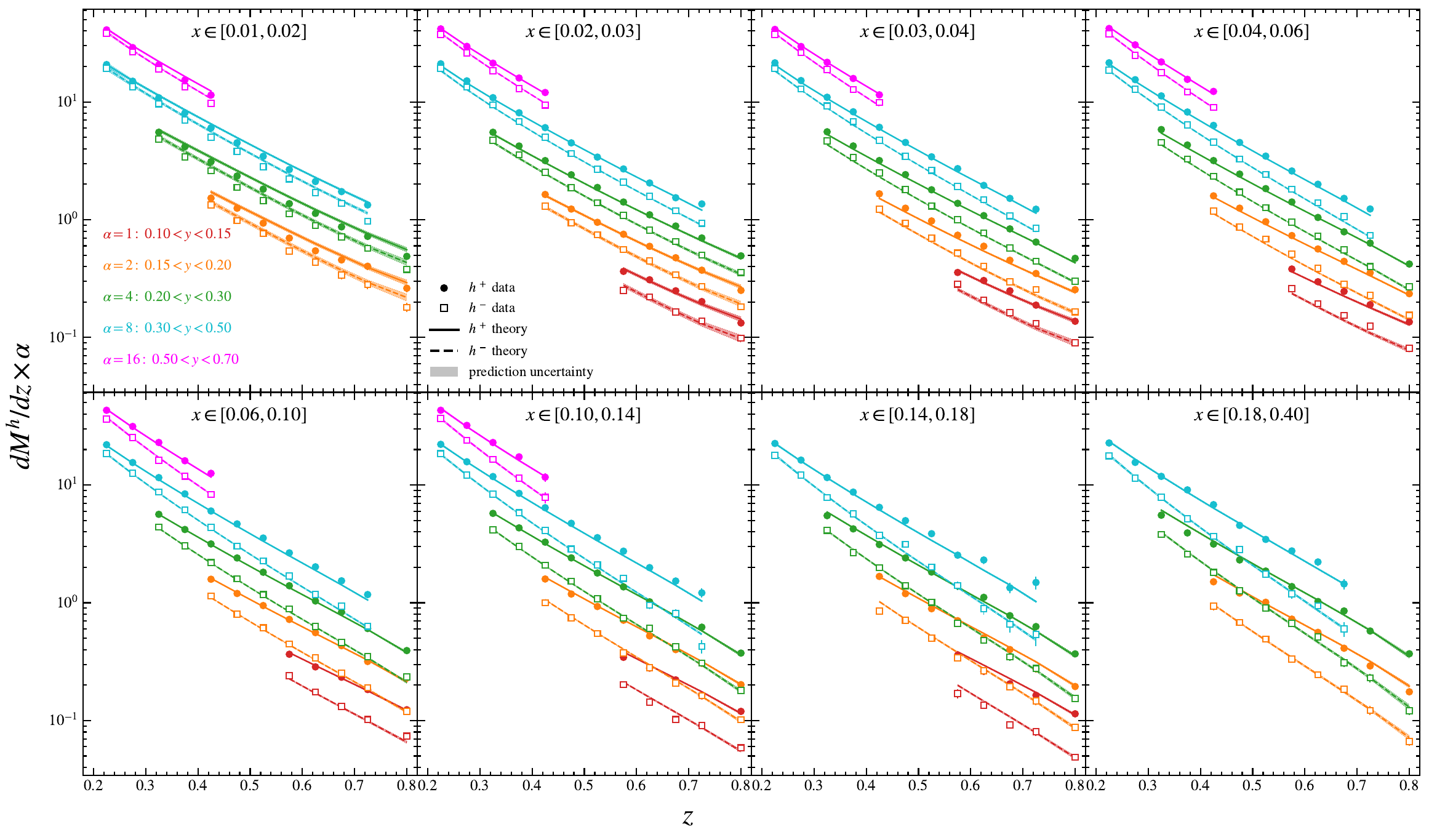}
\caption{\small
Revised multiplicities for unidentified charged hadrons
provided in the COMPASS 2026~\cite{COMPASS:2025bfn},
compared with the final fitted theory predictions. These multiplicities
supersede the original COMPASS 2017 isoscalar results. The same symbol,
line, and uncertainty conventions as in Fig.~\ref{fig:compass2025-hpm}
are used.
}
\label{fig:compass2026-hpm}
\end{figure*}

Figures~\ref{fig:compass2025-hpm} and~\ref{fig:compass2026-hpm}
show the comparison between the final fitted theory predictions and the
COMPASS multiplicities for unidentified charged-hadron production. The
COMPASS 2025 measurement is shown in
Fig.~\ref{fig:compass2025-hpm}, while the COMPASS 2026 are shown in
Fig.~\ref{fig:compass2026-hpm}. Both figures display the multiplicities
as functions of the hadron energy fraction $z$ in bins of $x$ and $y$.
Filled circles denote the $h^+$ data, open squares denote the $h^-$
data, solid curves show the fitted predictions for $h^+$, and dashed
curves show the corresponding predictions for $h^-$. The different $y$
intervals are separated by the multiplicative factors
$\alpha=1,2,4,8,16$, as indicated in the first panel. The vertical
bars represent the experimental uncorrelated uncertainties used in the
fit. The plotted predictions correspond to the final global fit, for
which the SIDIS data are selected with $Q_{\min}=1.4~\mathrm{GeV}$.

The shaded regions show the prediction uncertainty propagated from the
Monte Carlo replica ensemble to the SIDIS multiplicities. For each
kinematic point, the central curve corresponds to the fitted theory
prediction, while the band represents the one-standard-deviation spread
of the corresponding replica predictions. Since we follow the same
MontBlanc strategy, in which a PDF replica is assigned to each
fitted FF replica in the SIDIS calculation, this uncertainty includes
both the uncertainty of the fitted FF ensemble and the propagated PDF
uncertainty. It does not include additional uncertainties from variations
of $\alpha_s$, renormalization or factorization scales, or missing
higher-order corrections. 
To improve the visibility of the very narrow uncertainty bands, they are
displayed with darker shading and outlined by thin boundary curves. This
choice affects only the graphical representation; the numerical
uncertainties are not modified or enlarged.
The COMPASS 2025 proton-target multiplicities in
Fig.~\ref{fig:compass2025-hpm} shows that the fitted predictions describe
the main features of the measured $h^+$ and $h^-$ distributions. The
theory follows the overall decrease of the multiplicities with increasing
$z$ across the displayed $x$ and $y$ bins and reproduces the observed
charge separation between $h^+$ and $h^-$. In many bins the $h^+$
multiplicities are larger than the corresponding $h^-$ multiplicities,
reflecting the flavor sensitivity of the proton target and its enhanced
sensitivity to $u$-dominated PDF-weighted combinations. This feature is
particularly important for the present analysis, since the proton-target
data provide direct charge-separated information on the light-quark and
antiquark components of the unidentified charged-hadron FFs.

The agreement in Fig.~\ref{fig:compass2025-hpm} is generally good over
most of the measured kinematic range. The fitted curves describe both the
normalization and the shape of the data in most panels. Visible
deviations, where present, occur mainly near the edges of the measured
phase space, especially at larger values of $z$, in the highest-$x$ bins,
or in bins with fewer data points. These regions are more sensitive to
endpoint effects, reduced statistics, possible residual power corrections,
and limitations of the fixed-order leading-power description. The
prediction bands are narrow in most of the well-constrained
intermediate-$z$ region and are often comparable to, or smaller than, the
thickness of the central theory curves. They become visibly broader
mainly in the less constrained parts of the displayed kinematic range,
in particular at small $x$, small $y$, and toward larger $z$.

Figure~\ref{fig:compass2026-hpm} shows the corresponding comparison for
the multiplicities from the COMPASS 2026. These multiplicities supersede the original COMPASS 2017
isoscalar results and provide the updated isoscalar SIDIS input used in
the present analysis~\cite{COMPASS:2016xvm,COMPASS:2016crr,COMPASS:2025bfn}.
The fitted predictions reproduce the main $z$ dependence of both $h^+$
and $h^-$ multiplicities and give a good description of the charge
separation over a broad range of $x$ and $y$. Compared with the
proton-target data, the isoscalar target provides a more balanced
light-flavor weighting and therefore gives complementary information on
the quark and antiquark FFs. The successful description of the revised
isoscalar multiplicities provides an important consistency check of the
present analysis, since these data replace the older COMPASS isoscalar
input used in previous charged-hadron FF determinations.

The uncertainty representation is the same in
Figs.~\ref{fig:compass2025-hpm} and~\ref{fig:compass2026-hpm}. In both
cases, the experimental uncertainties correspond to the uncorrelated
uncertainties entering the diagonal part of the covariance matrix, while
the shaded prediction bands represent only the FF-replica uncertainty of
the fitted theory curves. The small size of the prediction bands in most
bins reflects the strong constraints provided by the combined SIA and
modern COMPASS SIDIS data in the displayed kinematic region, rather than
an estimate of the full theoretical uncertainty. The overall agreement
between the data points and the theory curves indicates that the final
fit captures the dominant normalization and shape features of the
measured multiplicities.

%The two COMPASS inputs play complementary roles in the global extraction.
%The proton-target data of Ref.~\cite{COMPASS:2024gje} enhance
%sensitivity to $u$-dominated flavor combinations, while the revised
%isoscalar-target multiplicities of Ref.~\cite{COMPASS:2025bfn}
%provide a more balanced constraint on the light-quark sector. Their
%simultaneous description with a common set of FFs supports the internal
%consistency of the modern COMPASS SIDIS input. Together, the
%proton-target and revised isoscalar-target multiplicities provide the
%charge-separated SIDIS information needed to improve the flavor
%decomposition of unidentified charged-hadron FFs. These comparisons
%therefore support the use of the modern COMPASS SIDIS multiplicities as
%the appropriate replacement for the older COMPASS input in a global
%charged-hadron FF analysis.

%---------------------------------------------------------------------
%---------------------------------------------------------------------

%---------------------------------------------------------------------
%---------------------------------------------------------------------

%%%%%%%%%%%%%%%%%%%%%%%%%%%%%%%%%%%%%%%%%%%%%%%%%%%%%%%%%%%%%%%%%%%%%%%
\subsection{The \texttt{HAPS-hFF1.0} fragmentation-function set}\label{subsec:nlo-nnlo-ffs}
%%%%%%%%%%%%%%%%%%%%%%%%%%%%%%%%%%%%%%%%%%%%%%%%%%%%%%%%%%%%%%%%%%%%%%%

\begin{figure*}[t]
\centering
\includegraphics[width=0.30\textwidth]{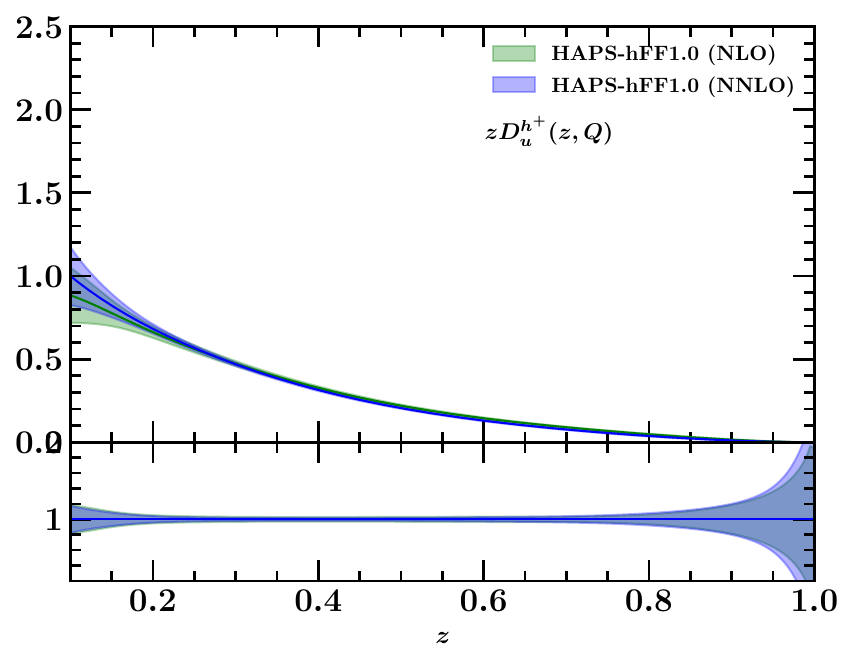}
\includegraphics[width=0.30\textwidth]{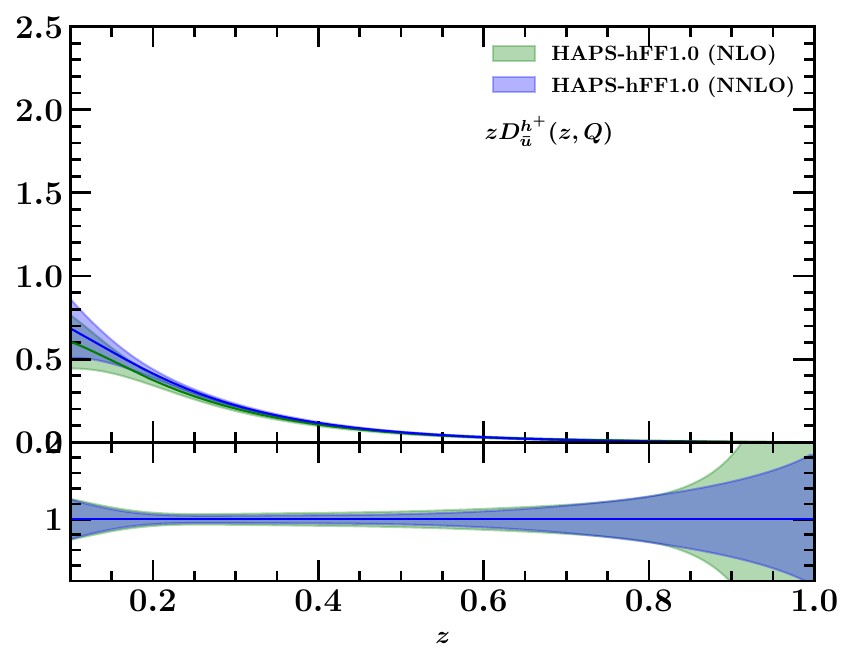}
\includegraphics[width=0.30\textwidth]{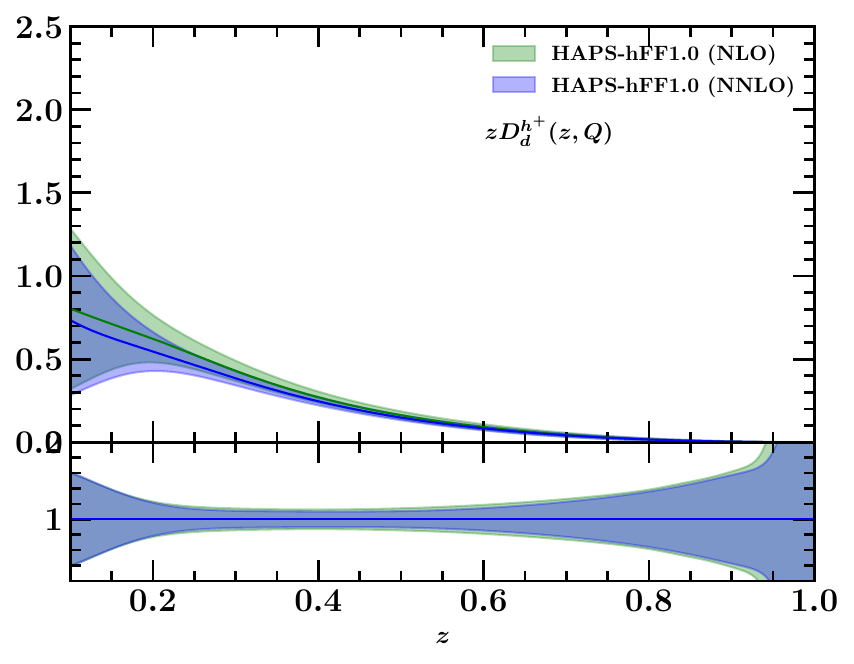}
\vspace{0.2cm}
\includegraphics[width=0.30\textwidth]{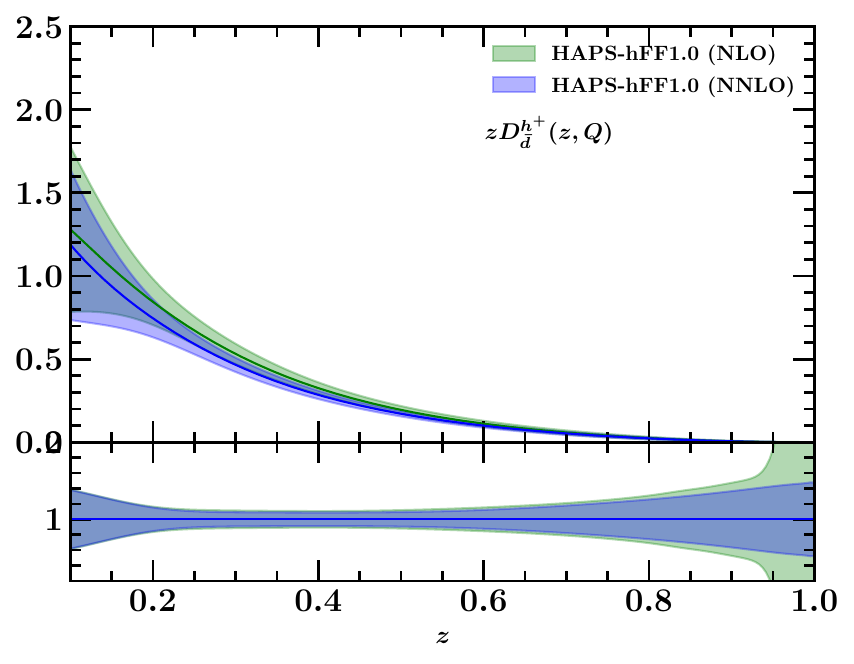}
\includegraphics[width=0.30\textwidth]{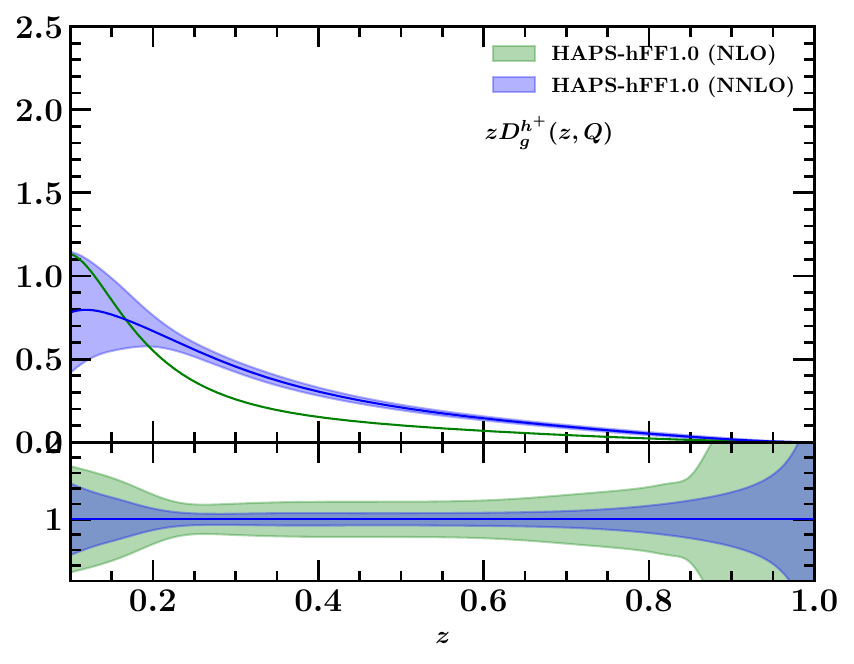}
\includegraphics[width=0.30\textwidth]{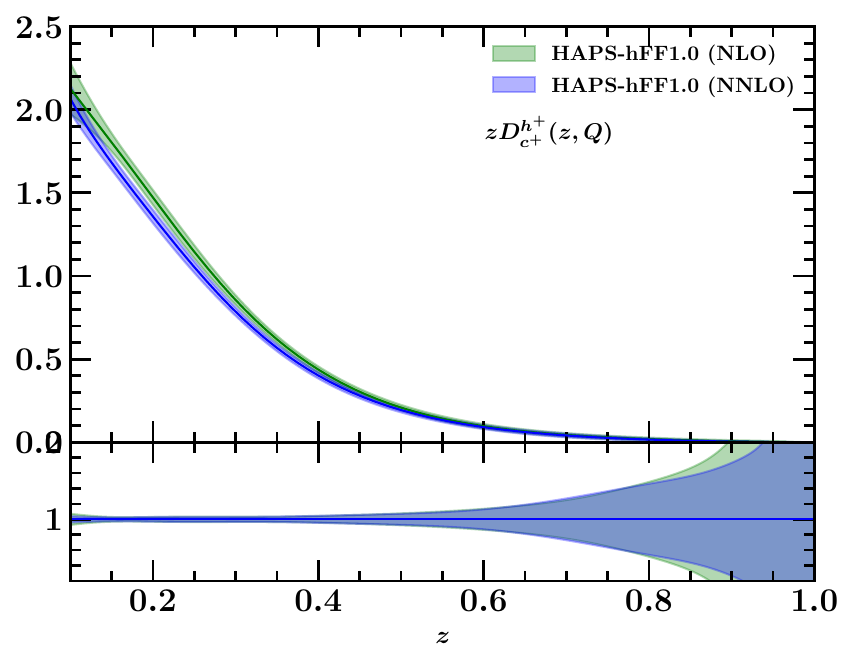}
\vspace{0.2cm}
\includegraphics[width=0.30\textwidth]{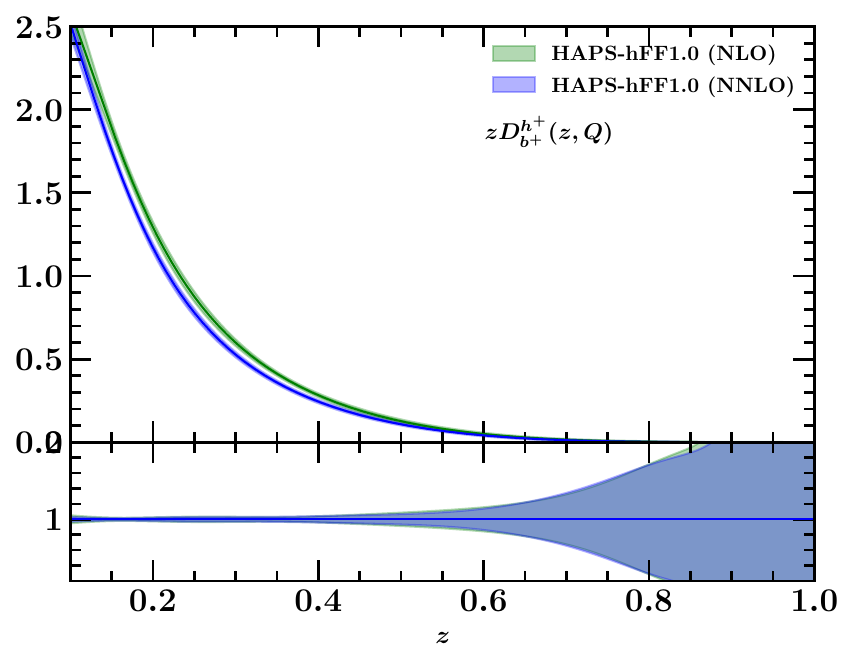}
\caption{\small
Comparison of the \texttt{HAPS-hFF1.0} unidentified charged-hadron FFs
at NLO and NNLO at \(Q=5~\mathrm{GeV}\). The panels show
\(zD_i^{h^+}(z,Q)\) for the independent flavor combinations
\(u,\bar u,d,\bar d,g,c^+\), and \(b^+\). Green bands denote the NLO
result and blue bands denote the NNLO result. The lower panel in each
plot shows the self-normalized ratio for each perturbative order,
obtained by dividing the corresponding NLO or NNLO replica band by its
own central prediction. The bands represent the FF uncertainty estimated
from the fitted replica ensemble. }
\label{fig:ff-nlo-nnlo}
\end{figure*}

We now present the resulting \texttt{HAPS-hFF1.0} determination of
unidentified charged-hadron fragmentation functions. 
The comparison between the NLO and NNLO extractions is shown in
Fig.~\ref{fig:ff-nlo-nnlo} for the independent unidentified charged-hadron
FFs at $Q=5~\mathrm{GeV}$
\[
D_u^{h^+},\quad
D_{\bar u}^{h^+},\quad
D_{d+s}^{h^+},\quad
D_{\bar d + \bar s}^{h^+},\quad
D_g^{h^+},\quad
D_{c^+}^{h^+},\quad
D_{b^+}^{h^+}.
\]
The upper panel of each plot displays \(zD_i^{h^+}(z,Q)\), while the
lower panel shows the corresponding self-normalized ratios, obtained by
dividing each NLO and NNLO replica band by its own central prediction.
The green bands correspond to the NLO extraction and the blue bands to
the NNLO extraction. 

Overall, the NLO and NNLO FFs are broadly compatible over most of the
fitted $z$ range. This agreement is particularly clear for the quark and
heavy-quark combinations, where the central values are close and the
uncertainty bands largely overlap. The $u$-quark FF shows very good
stability between the two perturbative orders, with only small shifts in
the central value. The $\bar u$ FF is also compatible between NLO and
NNLO, with somewhat larger differences appearing mainly near the edges of
the fitted $z$ range. The $d$ and $\bar d$ FFs display a similar pattern,
although their uncertainty bands are generally broader and their
order-by-order variation is more visible than in the $u$ sector. This is
expected because the $d$-type combinations are less directly constrained
by the proton-dominated SIDIS flavor weights.

The heavy-quark combinations, $D_{c^+}^{h^+}$ and $D_{b^+}^{h^+}$, show a
high degree of perturbative stability. Their NLO and NNLO central values
are close over the fitted range, and their uncertainty bands are
well controlled. This behavior reflects the dominant role of
flavor-tagged SIA data in constraining the heavy-quark FFs. 
By contrast, the gluon FF exhibits the largest visible
difference between the NLO and NNLO extractions and also carries the
largest uncertainty band. This is consistent with the fact that, in an
SIA+SIDIS analysis, the gluon is constrained mainly indirectly through
scaling violations, higher-order coefficient functions, and correlations
with the quark sector, rather than through a direct leading-order SIDIS
sensitivity.

The ratio panels provide a more detailed view of the relative uncertainty
of each extraction. Since each band is normalized to its own central
prediction, the ratio panels should not be interpreted as direct
NLO/NNLO ratios. Instead, they illustrate the size of the replica
uncertainties around the corresponding NLO and NNLO central values. For
most light-quark and heavy-quark FFs, the relative uncertainties remain
moderate in the region where the data provide the strongest constraints.
Larger relative uncertainties are observed toward small and large values
of \(z\), where extrapolation effects, small-\(z\) logarithmic
contributions, endpoint sensitivity, and reduced experimental constraints
become more important. The gluon band shows the largest relative
uncertainty, confirming that the gluon FF remains the least directly
constrained component of the present SIA+SIDIS extraction.

This behavior is consistent with the global fit qualities discussed
above. The NNLO fit gives a modestly smaller total value $\chi^2/N_{\rm dat}=0.873$ compared with
$\chi^2/N_{\rm dat}=0.943$ at NLO. 
The improvement in the total fit quality is reflected in the
slightly shifted NNLO central values, while the broad overlap of the NLO
and NNLO uncertainty bands indicates that the main features of the
extraction are stable. We do not interpret the smaller NNLO
$\chi^2/N_{\rm dat}$ as evidence that NNLO improves every individual
dataset or every flavor component. Rather, the comparison should be viewed
as a perturbative-stability test of the global extraction. 

It is also important to recall the theoretical status of the NNLO
calculation used here. 
The SIA coefficient functions and timelike DGLAP evolution are included
at NNLO accuracy, while the SIDIS contribution is treated within the
NNLO perturbative setup implemented in the present framework~\cite{AbdulKhalek:2022laj,MontBlancCode, MontBlancZenodo}. 
The NNLO FFs should therefore be interpreted in light of this theoretical setup.
Within this framework, Fig.~\ref{fig:ff-nlo-nnlo} shows that the
quark-sector FFs are robust under the change of perturbative order, the
heavy-quark FFs are stable, and the gluon remains the least directly
constrained component. The updated COMPASS SIDIS data thus provide
stable constraints on the light-quark flavor separation in the present
global analysis.

%---------------------------------------------------------------------
%---------------------------------------------------------------------

%%%%%%%%%%%%%%%%%%%%%%%%%%%%%%%%%%%%%%%%%%%%%%%%%%%%%%%%%%%%%%%%%%%%%%%
\subsection{Comparison of \texttt{HAPS-hFF1.0} with NNFF1.1h}\label{subsec:comparison-nnff}
%%%%%%%%%%%%%%%%%%%%%%%%%%%%%%%%%%%%%%%%%%%%%%%%%%%%%%%%%%%%%%%%%%%%%%%

\begin{figure*}[t]
\centering
\includegraphics[width=0.30\textwidth]{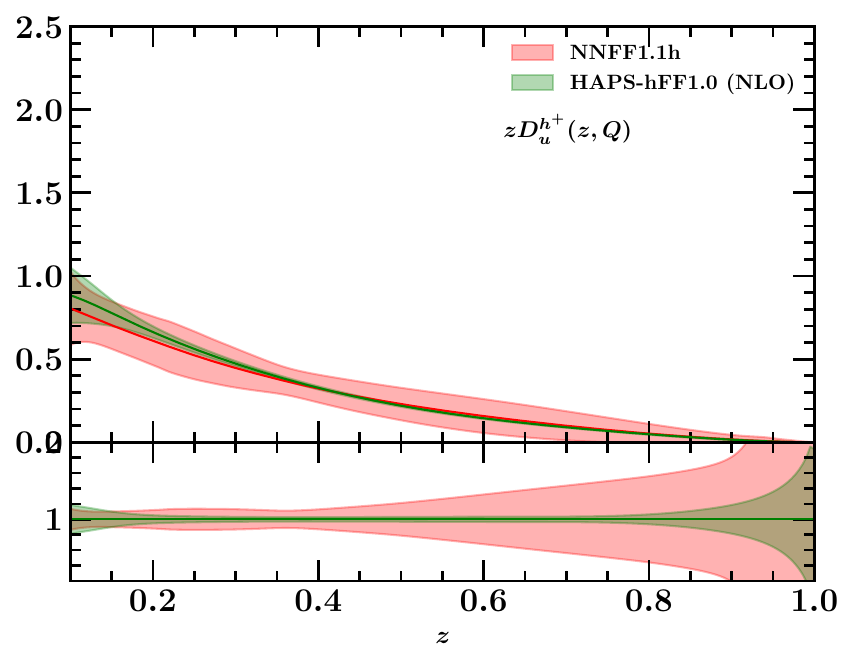}
\includegraphics[width=0.30\textwidth]{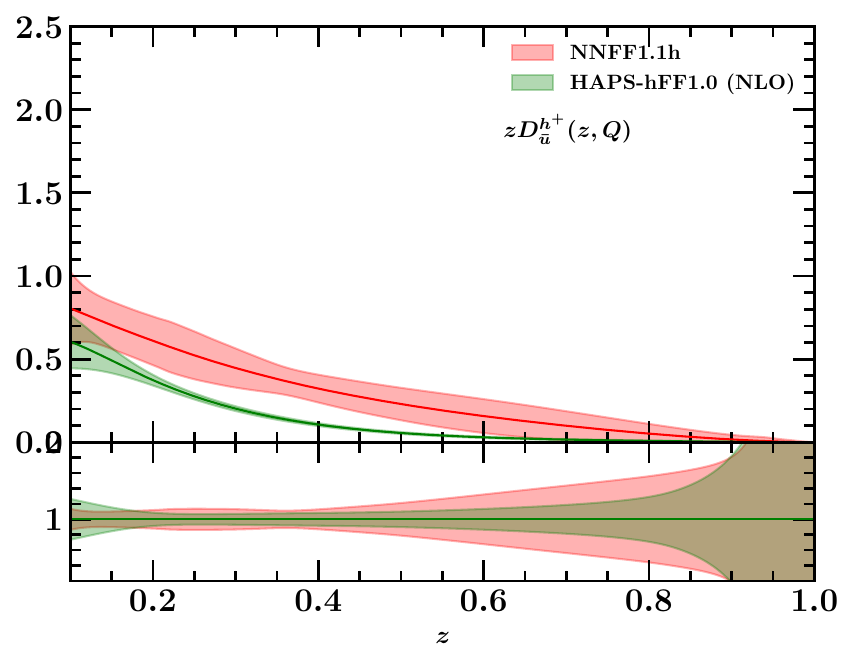}
\includegraphics[width=0.30\textwidth]{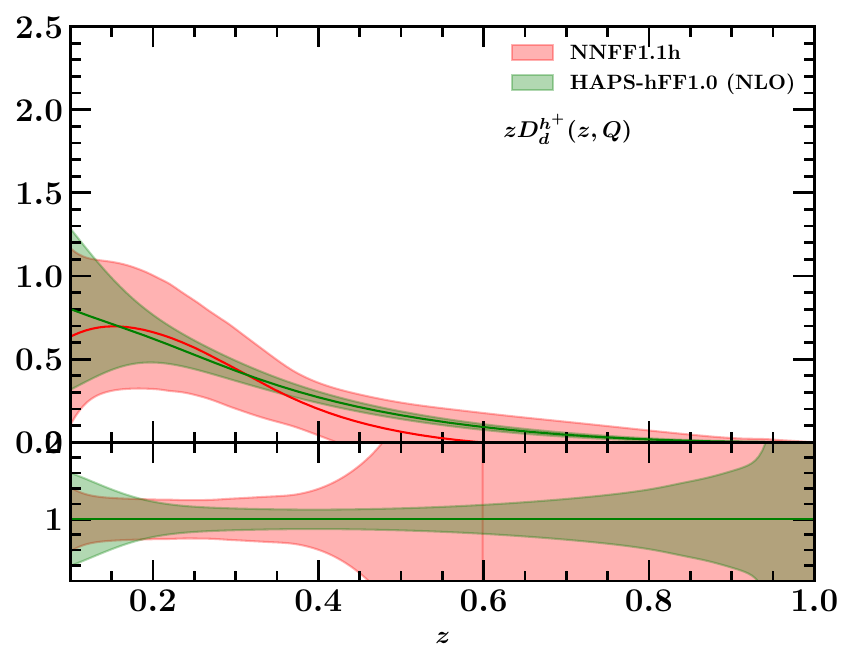}
\vspace{0.2cm}
\includegraphics[width=0.30\textwidth]{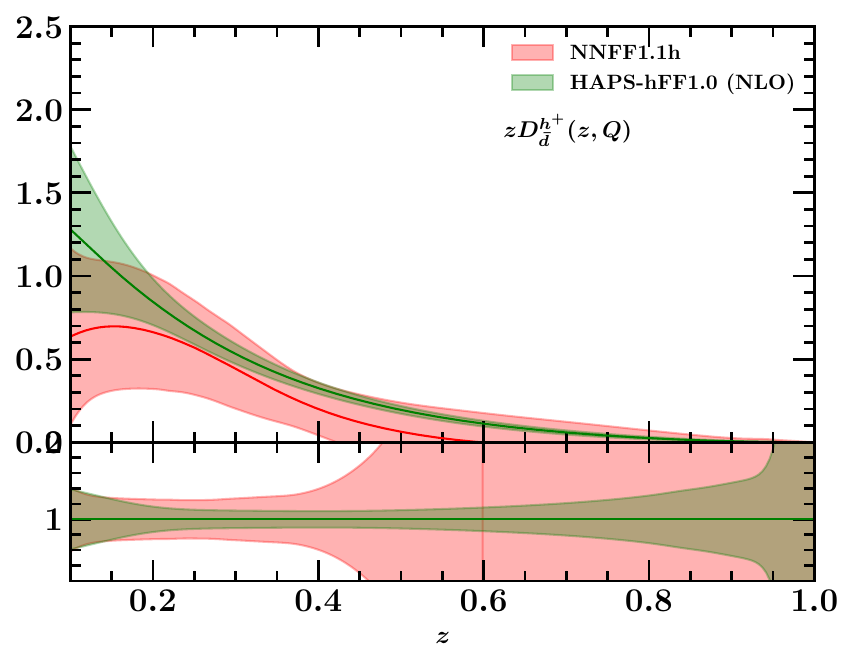}
\includegraphics[width=0.30\textwidth]{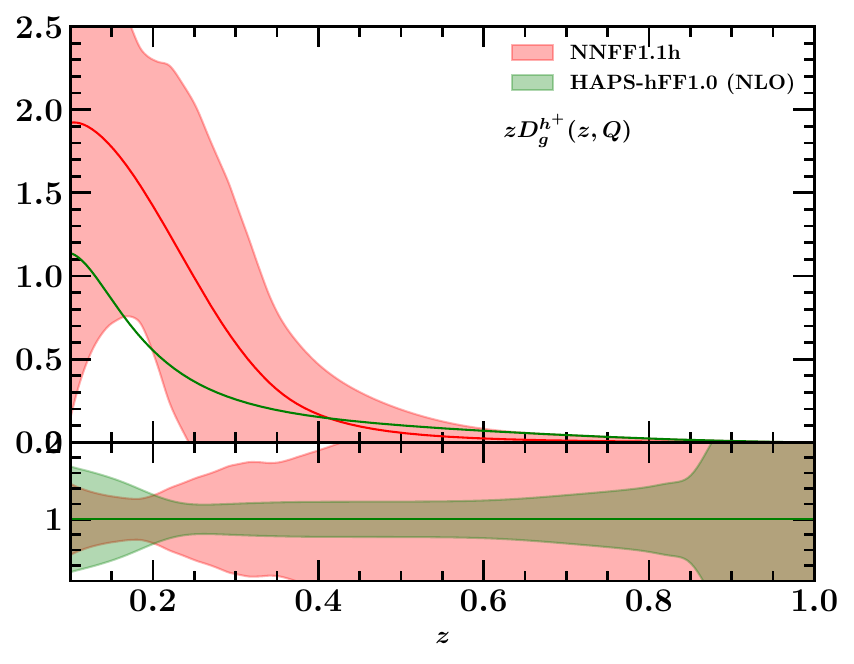}
\includegraphics[width=0.30\textwidth]{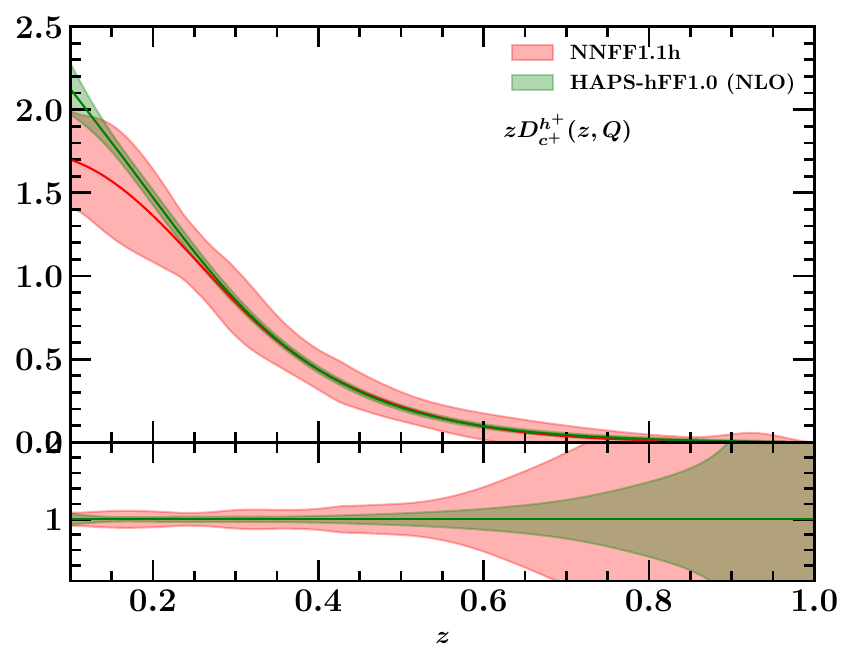}
\vspace{0.2cm}
\includegraphics[width=0.30\textwidth]{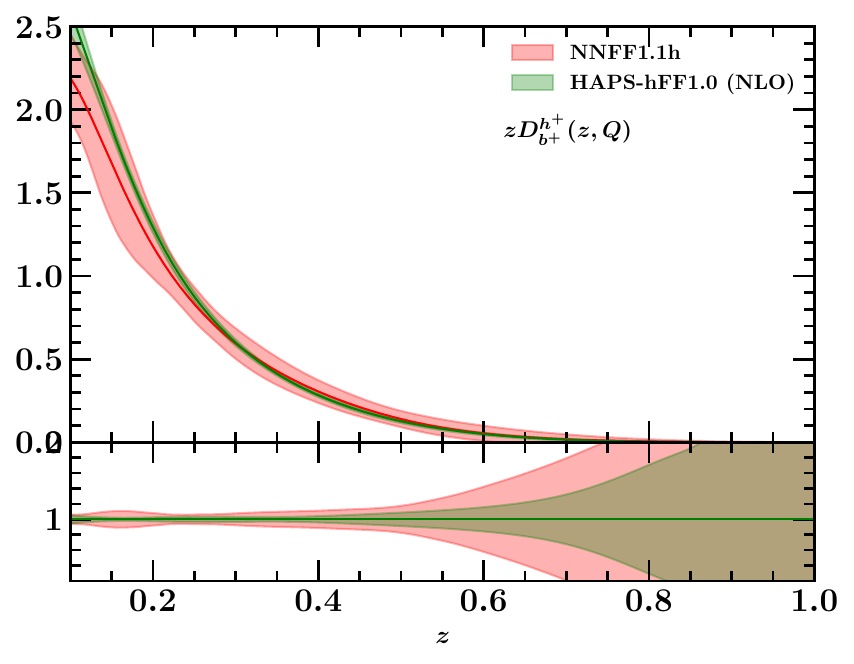}
\caption{\small
Comparison of the present \texttt{HAPS-hFF1.0} NLO FFs with
NNFF1.1h~\cite{Bertone:2018ecm} at \(Q=5~\mathrm{GeV}\). Red bands
denote NNFF1.1h and green bands denote the present
\texttt{HAPS-hFF1.0} NLO result. The panels show
\(zD_i^{h^+}(z,Q)\) for the independent flavor combinations, together with the corresponding relative uncertainty bands in the lower
panel of each plot, obtained by normalizing each FF ensemble to its own
central value. The comparison is performed at NLO because NNFF1.1h is an NLO
determination. The largest difference appears in the gluon FF, reflecting
the fact that NNFF1.1h includes hadron-collider data that directly
constrain gluon fragmentation, whereas the present analysis uses SIA and
modern COMPASS SIDIS multiplicities. }
\label{fig:ff-nnff}
\end{figure*}

In Fig.~\ref{fig:ff-nnff}, we compare the present
\texttt{HAPS-hFF1.0} NLO FFs with the NNFF1.1h
determination~\cite{Bertone:2018ecm} at $Q=5~\mathrm{GeV}$. The
comparison is shown for the independent unidentified charged-hadron FF
combinations
\[
D_u^{h^+},\quad
D_{\bar u}^{h^+},\quad
D_{d+s}^{h^+},\quad
D_{\bar d + \bar s}^{h^+},\quad
D_g^{h^+},\quad
D_{c^+}^{h^+},\quad
D_{b^+}^{h^+}.
\]
The upper panel of each plot displays $zD_i^{h^+}(z,Q)$, while the lower
panel shows the corresponding relative uncertainty bands, obtained by
normalizing each FF ensemble to its own central value. 
The red bands denote NNFF1.1h, and the green bands denote the present
\texttt{HAPS-hFF1.0} NLO result.

%---------------------------------------------------------------------
%---------------------------------------------------------------------

This comparison is useful for assessing the consistency of the present
extraction with an earlier determination of unidentified charged-hadron
FFs, but it should not be interpreted as a direct one-to-one comparison
of two identical analyses. The two fits are based on different
experimental inputs and different methodologies. The present
\texttt{HAPS-hFF1.0} analysis is a direct global refit using SIA data
together with the new COMPASS 2025 SIDIS multiplicities
and the revised COMPASS 2026 SIDIS multiplicities.
These charge-separated SIDIS data provide direct sensitivity to the
light-quark and antiquark flavor structure. In contrast, NNFF1.1h is an
NLO determination based on an SIA prior supplemented by hadron-collider
data through Bayesian reweighting~\cite{Bertone:2018ecm}. The collider
data included in NNFF1.1h provide strong sensitivity to the gluon and
singlet FF combinations, while the present analysis is designed to
isolate the impact of the updated COMPASS SIDIS input.

For the $u$-quark FF, the \texttt{HAPS-hFF1.0} and NNFF1.1h central
values are broadly compatible over much of the fitted $z$ range, with
overlapping uncertainty bands in the region where the data provide the
strongest constraints. The situation is somewhat different for the
$\bar u$ FF. In this channel, the present \texttt{HAPS-hFF1.0} result is
shifted downward with respect to NNFF1.1h and lies outside the NNFF1.1h
one-standard-deviation band in part of the fitted $z$ range. This
difference should not be interpreted as a direct inconsistency between
the two determinations, since the two fits use different data sets and
different constraining mechanisms. In the present analysis, the
charge-separated COMPASS SIDIS multiplicities provide direct information
on quark--antiquark separation, whereas NNFF1.1h is based on an SIA
prior supplemented by hadron-collider data through Bayesian reweighting.
The observed shift in the $\bar u$ channel therefore reflects the
additional flavor information entering the present SIA+SIDIS fit.

The $d$-type combinations also show visible differences. This is not
unexpected, because these channels are less directly constrained than the
$u$-dominated combinations in proton-target SIDIS, and their extraction
is more sensitive to the interplay between proton and isoscalar targets.
The differences in the low- and intermediate-$z$ regions therefore
reflect the different flavor information entering the two analyses. The
comparison should also be viewed in light of the fact that the present
fit uses the revised COMPASS data together with the new COMPASS data, while NNFF1.1h does not include these
charge-separated SIDIS multiplicities in the same way.

The largest difference between the two determinations appears in the
gluon FF. This is the most important distinction in
Fig.~\ref{fig:ff-nnff}. The gluon in NNFF1.1h is strongly affected by
the inclusion of hadron-collider charged-particle spectra, which provide
direct sensitivity to gluon fragmentation. The present
\texttt{HAPS-hFF1.0} analysis does not include hadron-collider data; in an
SIA+SIDIS fit, the gluon is constrained more indirectly through scaling
violations, higher-order corrections, and correlations with the quark
sector. The difference in the gluon FF should therefore not be
interpreted as an inconsistency between the two determinations. Rather,
it reflects the complementary constraining power of SIDIS and
hadron-collider observables.

The uncertainty band of the gluon FF should also be interpreted with
care. Since no proton--proton charged-hadron production data are included
in the present fit, the gluon is not directly constrained by
hadron-collider observables. The green band shown in
Fig.~\ref{fig:ff-nnff} represents the Monte Carlo replica uncertainty
within the adopted data set, parametrization, kinematic cuts, and
theoretical framework. It should not be regarded as a complete estimate
of all possible methodological and missing-theory uncertainties in the
gluon sector. In particular, residual dependence on the functional
flexibility, the data selection, the treatment of correlations, and the
absence of direct $pp$ constraints may still be relevant. The neural-network
parametrization reduces functional-form bias compared with more rigid
analytic forms, but it does not by itself eliminate all possible sources
of methodological uncertainty.

The heavy-quark combinations, $D_{c^+}^{h^+}$ and $D_{b^+}^{h^+}$, show
a more stable behavior. Their central values are broadly compatible with
NNFF1.1h, and the shapes are similar over most of the displayed $z$
range. This agreement is expected because the heavy-quark FFs are mainly
constrained by flavor-tagged SIA measurements, which are common in
spirit to both analyses and are less directly affected by the inclusion
of COMPASS SIDIS multiplicities.

The uncertainty bands in Fig.~\ref{fig:ff-nnff} therefore illustrate the
different sources of information in the two fits. In the light-quark and
antiquark sectors, the differences in central values show the impact of
the updated charge-separated COMPASS SIDIS data. In the gluon sector,
the difference is mainly driven by the absence of direct hadron-collider
constraints in the present fit. In the poorly constrained small- and
large-$z$ regions, the uncertainty bands widen and the ratio panels show
larger deviations from unity; these regions should therefore not be
overinterpreted.

The comparison in Fig.~\ref{fig:ff-nnff} is restricted to the NLO
determination, since NNFF1.1h is an NLO analysis~\cite{Bertone:2018ecm}.
A direct comparison of the \texttt{HAPS-hFF1.0} NNLO result with NNFF1.1h
is therefore not possible. More generally, previous NNLO studies of
unidentified charged-hadron FFs exist in the SIA-only
context~\cite{Soleymaninia:2018uiv}.  
\texttt{HAPS-hFF1.0} is the first determination of unidentified 
charged-hadron FFs incorporating the
new COMPASS 2025 data and the revised COMPASS 2026 data up to NNLO accuracy.  For this reason, external
comparisons with previous unidentified charged-hadron FF sets are made
at NLO, while the \texttt{HAPS-hFF1.0} NNLO result is used primarily to
assess the perturbative stability of the present SIA plus
updated-COMPASS SIDIS extraction. 

In summary, Fig.~\ref{fig:ff-nnff} shows that the present
\texttt{HAPS-hFF1.0} NLO set is broadly consistent with NNFF1.1h in
several quark and heavy-quark channels, while it also reveals notable
differences in the light-antiquark and gluon sectors. The shift observed
in the $\bar u$ FF reflects the impact of the updated charge-separated
COMPASS SIDIS information, while the gluon comparison highlights the
importance of future analyses combining SIA, updated COMPASS SIDIS, and
hadron-collider charged-particle data in a single framework in order to
obtain simultaneous light-flavor separation and a more direct gluon
constraint.

%%%%%%%%%%%%%%%%%%%%%%%%%%%%%%%%%%%%%%%%%%%%%%%%%%%%%%%%%%%%%%%%%%%%%%%
\section{Summary and Conclusions}\label{sec:Summary} 
%%%%%%%%%%%%%%%%%%%%%%%%%%%%%%%%%%%%%%%%%%%%%%%%%%%%%%%%%%%%%%%%%%%%%%%

We have presented \texttt{HAPS-hFF1.0}, a new global QCD analysis of
unidentified charged-hadron fragmentation functions based on SIA data and
the modern COMPASS SIDIS multiplicities. The analysis updates the SIDIS
input used in earlier charged-hadron FF determinations by incorporating
the COMPASS 2025 proton-target measurement together with the revised 
$^{6}\mathrm{LiD}$-target multiplicities provided in the 2026 COMPASS addendum.

The revised multiplicities supersede the original
COMPASS 2017 isoscalar results and therefore provide the appropriate
updated isoscalar input for a modern SIA+SIDIS extraction. Together with
the proton-target measurement, they form a consistent COMPASS SIDIS data
set for reassessing unidentified charged-hadron FFs. The proton-target
multiplicities provide enhanced charge-separated sensitivity to
\(u\)-dominated flavor combinations, while the revised
multiplicities provide a more balanced light-flavor constraint.

The modern COMPASS multiplicities are described consistently together
with the SIA data by a common set of FFs. Their inclusion provides
important constraints on the light-quark and antiquark fragmentation
functions and improves the flavor decomposition relative to analyses
based on the older COMPASS isoscalar input. This supports the use of the
modern COMPASS SIDIS multiplicities as the appropriate replacement for
the previous COMPASS input in future charged-hadron FF analyses.

The extraction has been performed at both NLO and NNLO. Within the
theoretical framework used here, the comparison between the two
perturbative orders indicates good stability in the light-quark and
heavy-quark sectors. The gluon FF remains less directly constrained in
the present SIA+SIDIS-only analysis, as expected. This interpretation is
also supported by the comparison with NNFF1.1h, where larger differences
in the gluon sector reflect the impact of hadron-collider data included
in NNFF1.1h but not in the present fit.

This work therefore provides the first SIA+SIDIS determination of
unidentified charged-hadron FFs incorporating the modern COMPASS
proton-target measurement and the revised isoscalar-target multiplicities
from the COMPASS 2026.  The NNLO result should be viewed as a
perturbative-stability extension of this updated SIA+SIDIS extraction,
while previous NNLO determinations of unidentified charged-hadron FFs
were performed in an SIA-only setting.
The resulting \texttt{HAPS-hFF1.0} replicas are provided in LHAPDF format~\cite{Buckley:2014ana} 
for phenomenological applications, with public grids, documentation, and
plotting/usage examples available through the accompanying GitHub
repository~\cite{HAPS-hFF10}.

Future work should extend the present analysis by including
proton-proton and LHC charged-hadron production data in order to improve
the gluon constraint. It will also be important to study the impact of
projected EIC SIDIS measurements on flavor separation and to quantify
possible hadron-mass, target-mass, higher-twist, and nuclear effects in
the low-\(Q\) SIDIS region. In particular, recent LHCb measurements of
charged-hadron distributions in charm- and beauty-tagged jets~\cite{LHCb:2025zmu} provide a 
promising additional constraint on heavy-flavor fragmentation and offer a
useful testing ground for future charged-hadron FF sets.

%%%%%%%%%%%%%%%%%%%%%%%%%%%%%%%%%%%%%%%%%%%%%%%%%%%%%%%%%%%%%%%%%%%%%%%
\begin{acknowledgments}

The authors are grateful to the members of the COMPASS for making
their experimental data available to us. This
work was supported by the School of Particles and Accelerators at the
Institute for Research in Fundamental Sciences (IPM). Hamzeh Khanpour
appreciates the financial support from the IDUB program at AGH University of Kraków. 
Hubert Spiesberger acknowledges support by the 
Cluster of Excellence ``Precision Physics, Fundamental Interactions, 
and Structure of Matter" (PRISMA$^{++}$ EXC 2118/2) funded by the 
German Research Foundation (DFG) within the German Excellence 
Strategy (Project ID 390831469). 

\end{acknowledgments}
%%%%%%%%%%%%%%%%%%%%%%%%%%%%%%%%%%%%%%%%%%%%%%%%%%%%%%%%%%%%%%%%%%%%%%%

\newpage

%%%%%%%%%%%%%%%%%%%%%%%%%%%%%%%%%%%%%%%%%%%%%%%%%%%%%%%%%%%%%%%%%%%%%%%

\end{document}